\documentstyle[aps,psfig]{revtex}

\begin{document}

\title{Ground State Vortex Lattice Structures in $d$-wave
Superconductors}
\author{Sudhansu S. Mandal\cite{add1}\cite{add2} and T. V.
Ramakrishnan\cite{add1}}
\address{Centre for Condensed Matter Theory, Department of
Physics,
  Indian Institute of Science, Bangalore 560 012, India}

\date{\today}

\maketitle

\begin{abstract}
 We show in a realistic $d_{x^{2}-y^{2}}$ symmetry gap model for a cuprate
superconductor that the clean vortex lattice has discontinuous structural
transitions (at and near T=0), as a function of the magnetic field $B$ along
the $c$-axis.  The transitions arise from the singular nonlocal and anisotropic
susceptibility of the $d_{x^{2}-y^{2}}$ superconductor to the perturbation
caused by supercurrents associated with vortices.  The susceptibility, due to
virtual Dirac quasiparticle-hole excitation, is calculated carefully, and
leads to a ground state transition for the triangular lattice from an
orientation along one of the crystal axis to one at 45$^o$  to them, i.e,
along the gap zero direction. The field scale is seen to be $5$ Tesla $ \sim
(\Delta_{0}/ta)^{2}\Phi_{0}$, where $\Delta_{0}$ is the gap maximum, $t$ is
the nearest neighbour hopping, $a$ is the lattice constant, and $\Phi_{0}$ is
the flux quantum.  At much higher fields ($\sim 28T$) there is a discontinuous
transition to a centred square structure.  The source of the differences from
existing calculations, and experimental observability are discussed, the
latter especially in view of the very small (a few degrees $K$ per vortex)
differences in the ground state energy.  
\end{abstract}


\section{Introduction}

  An external magnetic field enters a (type II) superconductor as
a collection
of quantized magnetic flux tubes. The flux tubes with associated 
supercurrent vortices  form  a
triangular
lattice \cite{abrikosov} as is seen in conventional
superconductors 
\cite{essmann,hess}. Deviations from
this structure are of considerable interest. In conventional 
superconductors, the observed deviations have been attributed to 
the anisotropy of the underlying one electron energy spectrum.
In heavy fermion and high T$_c$ superconductors, an additional and
potentially 
very interesting reason is the existence of an unconventional 
superconducting order parameter, with gaps which have nodes and
change
sign. Indeed there is experimental evidence both in heavy fermion
systems \cite{h_fermion} the exotic
superconductor SrRuO$_4$ \cite{riseman} and in cuprate
superconductors \cite{maprile,keimer}that the vortex lattice is not
triangular (for some field and temperature regimes). Recently,
measurements of small
angle neutron diffraction from untwinned
YBa$_2$Cu$_3$O$_{7-\delta}$
single crystals shows a very well formed triangular lattice that
undergoes an
orientational transition from along $a$ axis to along $b$ axis for a $3T$
magnetic field at $33^o$ to the $c$-axis\cite{johnson}.  The reasons for
possible nontriangular structure as well as for the structural transition are
not fully established, and are the subject of considerable theoretical work
\cite{ichioka,maki,affleck,franz,amin}.
 In high T$_c$ superconductors, entropic effects, abetted by high
transition temperature as well as weak interlayer coupling play an important
role, and the classical statistical mechanics of interacting, meandering flux
lines, of the vortex fluid phase, and the solid fluid transition has developed
into a major  theoretical and experimental subfield \cite{blatter}.

In this paper, we focus on the nature of the flux lattice at
temperatures 
well below the superconducting transition, where vortex
configurational entropy 
effects mentioned above are negligible. The ground state
structures and 
structural transition then directly reflect the electronic
peculiarities of
the superconductor, and thus probe the latter. For cuprate
superconductors,
a number of measurements show that the superconducting gap
$\Delta_{\bf k}$
has nodes \cite{expts}, has a magnitude with a  $(\Delta_0/2) \vert
\cos k_xa -\cos k_ya \vert$ 
dependence on the
two dimensional wave vector ${\bf k}$ across the Fermi surface
\cite{norman2} and that
transport lifetimes of quasiparticles are long \cite{bonn} for
$T\ll T_c$. Thus
one can assume well defined low energy, nodal quasiparticles, with an
experimentally determined
one electron dispersion $\epsilon_{{\bf k}}$ \cite{norman1} and gap
function 
$\Delta_{{\bf k}}$ \cite{norman2}. The question of interest is the
effect of the zero
gap, anisotropic Dirac like linear quasiparticle excitation spectrum on the
interaction between vortices, and thus on vortex lattice
structure. There
is considerable evidence e.g., from magnetic field dependent
electronic
specific heat \cite{moler}, electronic thermal conductivity
\cite{krishana}, 
and superfluid density \cite{sonier}
measurements that an external magnetic field going in as vortices has a strong
effect
on electronic states, changing their density and lifetime. The
relevant
issue here is somewhat the reverse, namely the effect of the
quasiparticles
on interaction between vortices. The order parameter phase
associated with
the vortex, and the related magnetic vector potential together
constitute
the superfluid velocity field ${\bf v}_{s}({\bf r}) [=\sum_l {\bf v}_s ({\bf
r}-{\bf R}_l)]$ where the 
vortices are located at points ${\bf R}_l$. The extra superfluid
kinetic
energy, being quadratic in $v_s$, clearly has a part that depends
on two
vortex coordinates and is thus structure sensitive. In addition
to this
`diamagnetic' term, which is the additional kinetic energy of the
rigidly
moving superfluid and which is minimized for a triangular
lattice [1], there is
another `paramagnetic' term due to the perturbation of
quasiparticles by the 
superfluid current via the term ${\bf k}\cdot {\bf v}_{s}$. This causes virtual
particle hole excitations; two
vortices interact via the exchange of quasiparticle quasihole pairs. This
polarization
term depends on the quasiparticle excitation spectrum. In
a clean
$s$-wave superconductor, the process leads to an additional
isotropic
interaction between vortices of order $(H/H_{c_2})$ relative to
the diamagnetic
term. However, in a $d$-wave superconductor where the excitation
gap vanishes 
along some (nodal) directions, one expects the nonlocal polarizeability to be
larger as well as anisotropic;
this gives rise to an interaction between vortices which depends
on the 
orientation of the line joining them with the respect to crystaline axes,
and 
consequently can be the cause of novel long range positional
order.

The ground state energy arising from quasiparticle / hole
mediated
interaction between vortices depends linearly on the nonlocal
current 
susceptibility $\chi^{{\rm p}}_{\alpha\beta} ({\bf q})$, for wave vectors
${\bf q}$ 
equal to the reciprocal lattice vectors ${\bf G}$ of the vortex
lattice.
Because the gap as well as density of quasiparticle states
vanish
linearly near the node, $\chi^{{\rm p}}_{\alpha\beta} ({\bf q})$ is
proportional to
$\vert q_x \vert$ or $\vert q_y \vert$ for small $q$. This nonanalytic behaviour,
noticed 
first by Kosztin and Leggett \cite{leggett}, 
has also been discussed by Franz et al \cite{franz}
who were the first to analyze microscopically its effect, as well
as of the
anisotropy in $\chi^{{\rm p}}_{\alpha\beta}$, on vortex lattice structure
at $T=0$.
These authors found a rich phase diagram in the field temperature
plane,
with a centered  rectangular lattice at $T=0$ whose inner angle
varies
{\em continuously} as a function of field, as well as a sudden
orientational transition
at higher temperatures, and a transition to a centered square
lattice
for very high fields and low $T$. In obtaining these results,
Franz et al
made a ``local" approximation for the gap function, i.e., assumed
$\Delta_{{\bf k}}
=\Delta_{{\bf k}+{\bf G}}$, and more importantly they assumed a
momentum
independent quasiparticle current which leads to a
response function $\chi^{{\rm p}}_{xx}
({\bf G})=\chi^{{\rm p}}_{yy}({\bf G})$. The anisotropy then enters only through
$\chi^{{\rm p}}_{xy} ({\bf G})$. We
carry out here a more detailed and realistic calculation of the
nonlocal
susceptibility, considering the strong ${\bf k}$ dependence of
the gap function
$\Delta_{{\bf k}}$ and quasiparticle current ${\bf j}_{{\bf k}}$ properly, and
using a realistic 
one electron dispersion.
The diagonal terms $\chi^{{\rm p}}_{xx}$ and $\chi^{{\rm p}}_{yy}$ are unequal and
large, and
this anisotropy is seen to be the underlying cause of the
transition. 
The contibution of the off diagonal susceptitiblity $\chi^{{\rm p}}_{xy}$ is 
smaller than that of the diagonal susceptibilities.
Our results for structural
stability (at $T=0$) are therefore quite different from those of
Franz et al \cite{franz,amin}.

\begin{figure}[tb]
\psfig{figure=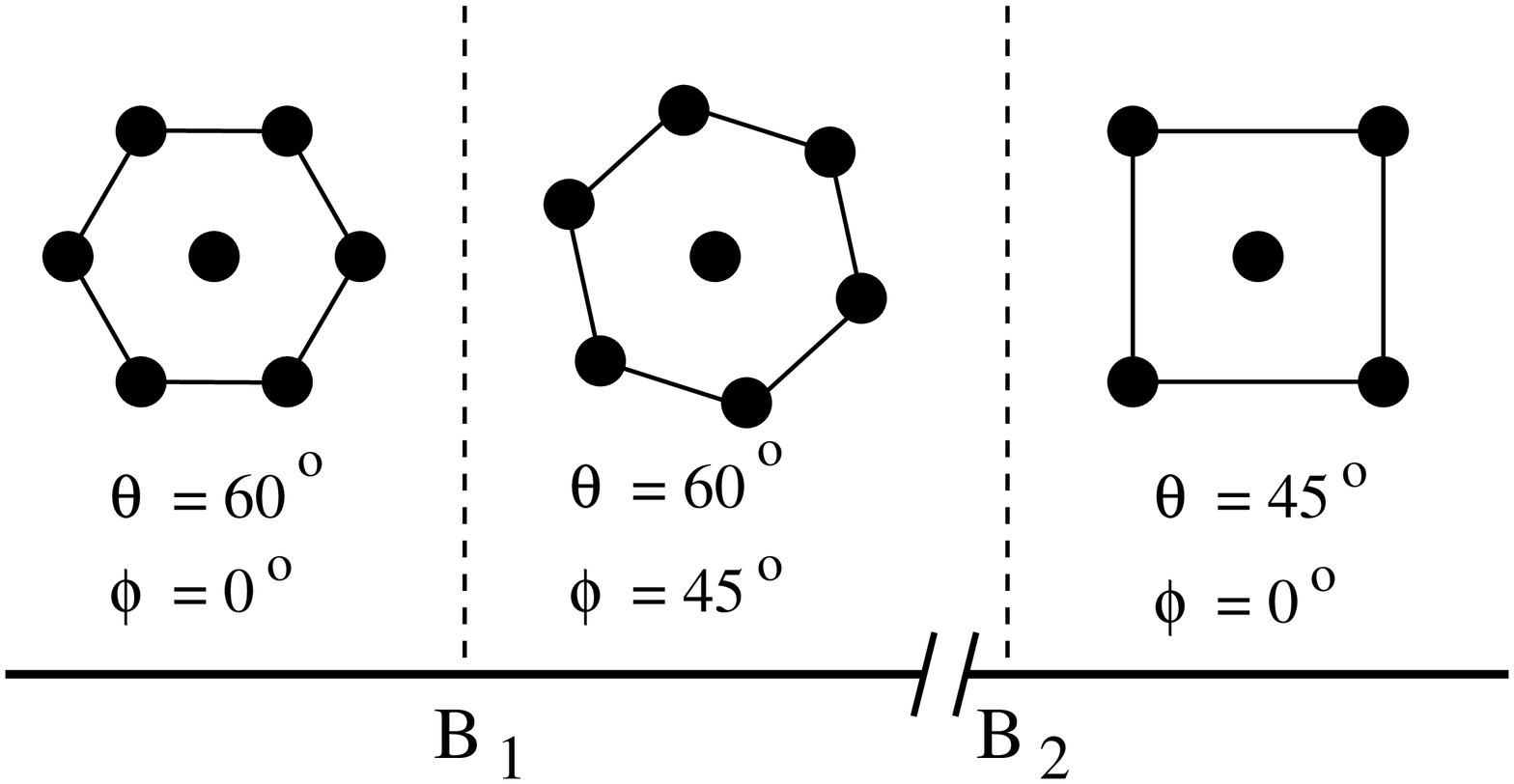,width=12cm,angle=-90}
\caption{Phase diagram for the structure of the vortex lattice.
The position of the vortices are denoted by black filled circles.
$B_1$ and $B_2$ are the fields at which the structural
transitions
take place as described in the text. The structures are shown 
diagrammatically.}
\label{phase_n}

\end{figure}

Confining ourselves to $T=0$, we find, as summarized in a phase
diagram
(see Fig.~\ref{phase_n}), 
that the stable lattice at low fields is triangular. At
about 5 Tesla (for the parameters chosen) the orientation of
the smallest ${\bf G}$ vector changes from being along one of the
axes to
being along the order parameter node direction, because the
system is most
susceptible to excitations with wave vector along the node. We have
analyzed the 
driving force for this transition, both analytically and
numerically, and
find that it arises from a subtle balance between the term linear
in $\vert
{\bf G} \vert$ and the quadratic term, which are slightly
different
for the two orientations. The field scale for the transition is 
approximately given by the condition $(taG/\Delta_0)\sim 1$ which
is natural
on dimensional grounds. The Fermi velocity is $ta$ and the energy
scale
associated with the superflow $({\bf \nabla}\theta)$ with Fourier
component $G$ 
is therefore $taG$. The polarizability or susceptibility has an energy
scale $(1/
\Delta_0)$ where $\Delta_0$ is the gap magnitude which sets the
scale for
quasiparticle excitation energies. Thus the dimensionless
susceptibility
of interest is $(taG/\Delta_0)$. For realistic parameters $t,a$,
and
$\Delta_0$ this translates $[(taG_c/\Delta_0)\simeq 0.37]$ to a
field scale
of 5.2T.

 We find that the node oriented triangular lattice is stable till
about 28T,
whereupon a discontinuous transition to a centered square lattice
takes place.
This structure, which is orientationally commensurate with the
symmetry
of the quasiparticle dispersion, is probably the most stable
$T=0$ phase when
electronic commensurate effects dominate. However, the calculated
field scale
is large enough that the London approximation used, valid for
$H\ll H_{c_2}$,
is not reliable, because vortex core effects cannot be neglected at these high
fields.

In the next section (Section II), we describe the model and the
theoretical
approach used. The tight binding quasiparticle Hamiltonian is
decomposed into
an unperturbed part $H_0$ and a term $H_I$ due to the
quasiparticle vortex
interaction. The free energy or the ground state energy can be
obtained as a 
power series in $H_I$ or equivalently the density of vortices.
For low vortex
densities $(H\ll H_{c_2})$ the leading or $n_v^2$ term is
sufficient, and
describes quasiparticle-hole mediated vortex interaction, in
addition to
the bare superfluid kinetic energy. We discuss the former
carefully in
terms of the nonlocal, anisotropic current susceptibility
$\chi^{{\rm p}}_{\alpha\beta}
({\bf q})$ since the energy can be expressed as a reciprocal
lattice vector sum
over $\chi^{{\rm p}}_{\alpha\beta}({\bf G})$. We obtain $\chi^{{\rm p}}_{\alpha\beta}
({\bf q})$ semianalytically for small $q$ at $T=0$, as well as numerically
(Section III).

\begin{figure}[tb]
\psfig{figure=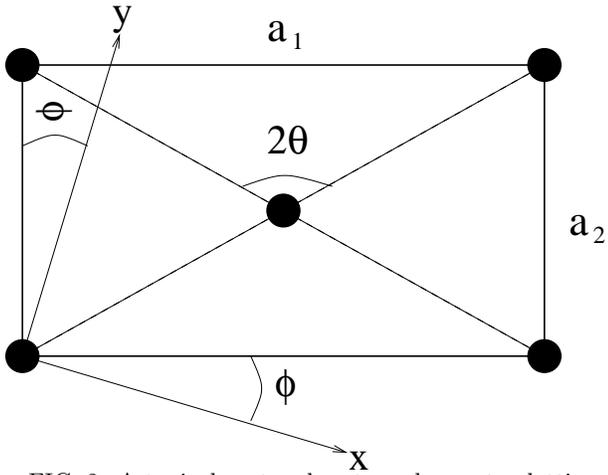,width=8cm,angle=-90}
\caption{A typical centered rectangular vortex lattice. Filled
circles represent
 the position of the vortices in the lattice. The aspect ratio of
the lattice is
 given by $a_1/a_2=\tan \theta$. The angle $\phi$ represents the
inclination of the
 lattice with respect to the crystal axis which are parallel to
$x$ and $y$ directions.}
\label{lat_n}
\end{figure}

 The calculations for different two dimensional structures are
discussed in                  
section III. For a given magnetic field $B$ the most general
centered 
rectangular lattice [$a_{1},a_{2}$] can be described in terms of an
angle $\theta$ related 
to the aspect ratio $(a_1/a_2)$ as $\tan \theta=a_1/a_2$, 
and an orientation $\phi$ 
with respect to crystal axes (Fig.~\ref{lat_n}). We compute the ground
state energy
as a 
function of these two variables for different magnetic fields.
The basic vortex related electronic energy parameters are the
following. The vortex
has three
energy scales, namely the diamagnetic single vortex energy, of
order 3450$K$
per vortex, the vortex interaction energy,  the diamagnetic
part of which has a value $\sim 1440K$ (for nearest neighbor
vortices), and the paramagnetic
interaction term which is about $350K$ at 5T field. 
The last, and the smallest term is structure sensitive and
is of interest here. The ground state is analyzed as a function
of $\theta$,
$\phi$ for several field values in section III. It turns out that
the
structure sensitive part of the last (paramagnetic vortex
interaction) term is extremely small, of
order a few
degrees per vortex. This has obvious implications for the 
observability of the transition, because the structural changes
predicted and the clean limit anisotropies obtained can be easily
overwhelmed by effects of disorder, eg., vortex pinning and the
muting of the paramagnetic susceptibility anisotropy, and
nonanalyticity by disorder.  However, the size of the structure
sensitive terms is larger, the greater the $(v_{F}/v_{\Delta})$
ratio or anisotropy. One can thus imagine
situations where this effect is quite large.

In the concluding section (Section IV) we briefly discuss thermal effects,
the consequence of the predicted transition and their observability,
the reason for which our result doffers from earlier, and the experimentally
observed structural transitions.

\section{Theory}

\subsection{Model}

 We consider a two dimensional
lattice model with nearest neighbor and next nearest neighbor
hopping for a CuO$_2$ plane of high $T_c$ superconductors. 
The (mean field) Hamiltonian in this model is given by
\begin{equation}
H_0 = -t\sum_{<ij>_{1}\sigma}(c_{i\sigma}^\dagger c_{j\sigma}+h.c.)
+t'\sum_{<<ij>>_{1}\sigma}(c_{i\sigma}^\dagger c_{j\sigma}+h.c.)
+\sum_{<ij>}(\Delta_{ij}c_{i\uparrow}^\dagger
c_{j\downarrow}^\dagger
+h.c.) -\mu\sum_i c_{i\sigma}^\dagger c_{i\sigma} \; ,
\label{eq1}
\end{equation}
where $t$ and $t'$ are nearest
neighbor and next nearest neighbor hopping integrals
respectively. 
This corresponds for appropriate choices of $t$ and $t'$ to an
open Fermi surface which is observed in angle resolved photo
emission
experiments. The pair amplitude 
$\Delta_{ij}$ is considered to be $d_{x^2-y^2}$-wave like, 
i.e., $\Delta_{i,i\pm a\hat{x}}
=-\Delta_{i,i\pm a\hat{y}}$, where $a$ is the lattice constant in
a square lattice, and $\mu$ is the chemical potential.

 When we apply a magnetic field beyond the lower critical field
$H_{c_1}$
in high $T_c$ superconductors, the magnetic  
field goes into the system in the form of vortices. The magnetic
induction is 
screened over a length $\lambda$, the penetration depth. 
 The pair amplitude
acquires a phase: $\Delta_{ij} \rightarrow \Delta_{ij}\exp
[-i\theta_{ij}]$,
where $\theta_{ij}$ is the sum of polar angles of all the vortices measured
with respect to a particular axis, for the centre of mass of the pair $ij$. We
write $\theta_{ij}$ as $(\theta_i +\theta_j)/2$, (as an average
of the angles of individual coordinates of the Cooper pairs), 
which is consistent upto ${\cal O}(1/k_F\xi)^{2}$, where $k_F$ is the
Fermi
momentum and $\xi$ is the superconductive coherence length. 
The pair amplitude (order parameter magnitude) vanishes at the center of 
the core of a 
vortex, and over a distance $\xi$ it acquires its uniform value.
For a collection of vortices with $H\ll H_{c_2}$, i.e., with intervortex
spacing $\gg \xi$, or the London limit, we assume the order parameter magnitude
be uniform, throughout the superconductor (there are $\delta$ function sources
of phase rotation at the locations of vortices). There is a vector potential
${\bf A}$ such that ${\bf \nabla} \times {\bf A}({\bf r})=B({\bf r})$
where $B({\bf r})$ is the local magnetic induction, along the $c$-axis.  Its effect
in this model is to change the hopping integrals to 
\begin{equation}
 (t,\, t' ) \rightarrow (t,\, t' )\, 
\exp [i(e/\hbar c)\int_{{\bf r}_i}^{{\bf r}_j} {\bf A}\cdot d{\bf
l} ]
\label{eq2}
\end{equation}
for hopping from site $j$ to site $i$.
We then make a gauge transformation: $c_{i\sigma}
\rightarrow c_{i\sigma}e^{-i\theta_i/2}$. We thus obtain the
Hamiltonian as
\begin{eqnarray}
H &=& -t\sum_{<ij>_{1}\sigma}(c_{i\sigma}^\dagger c_{j\sigma}\, \exp [
i(\theta_i-\theta_j)/2 +i(e/\hbar c)\int_{{\bf r}_i}^{{\bf
r}_j}{\bf A}\cdot d{\bf l}] +h.c.)
  +\sum_{<ij>_{1}\sigma}(\Delta_{ij}c_{i\uparrow}^\dagger
c_{j\downarrow}^\dagger +h.c.)
\nonumber \\
& & +t' \sum_{<<ij>>}(c_{i\sigma}^\dagger c_{j\sigma}\, 
\exp[ i(\theta_i-\theta_j)/2 +i(e/\hbar c)
\int_{{\bf r}_i}^{{\bf r}_j}{\bf A}\cdot d{\bf l}] +h.c.)
-\mu \sum_i c_{i\sigma}^\dagger c_{i\sigma}
\; .  \label{eq3}
\end{eqnarray}
The phase difference between two nearest or next nearest neighbour sites can be
expressed as \begin{equation}
 \frac{1}{2}(\theta_i-\theta_j)+(e/\hbar c)\int_{{\bf r}_i}^{{\bf
r}_j}
 {\bf A}\cdot d{\bf l}
\simeq
({\bf r}_i-{\bf r}_j)\cdot ( \frac{1}{2}{\bf \nabla}_i \theta
-(e/\hbar c){\bf A}_i)
\equiv (m/\hbar) ({\bf r}_i-{\bf r}_j)\cdot {\bf v}_s ({\bf r}_i)
\; . \label{eq4}
\end{equation}
Here the superfluid velocity 
\begin{equation}
{\bf v}_s({\bf r}) =
\frac{1}{m}\left[ \frac{\hbar}{2}\nabla \theta -\frac{e}{c} {\bf
A}({\bf r})\right]
\label{eq5}
\end{equation}
for a single vortex and for a collection of vortices, ${\bf v}_s
({\bf r})
= \sum_l {\bf v}_s({\bf r}-{\bf R}_l)$ (where the vortices are located at
${\bf R}_{l}$. We then assume that the phase difference between two
neighboring  lattice sites
is very small (which is certainly true in the London limit) so that we expand
exponentials in Eq.~(\ref{eq3}) upto
quadratic terms.

\begin{figure}[tb]
\psfig{figure=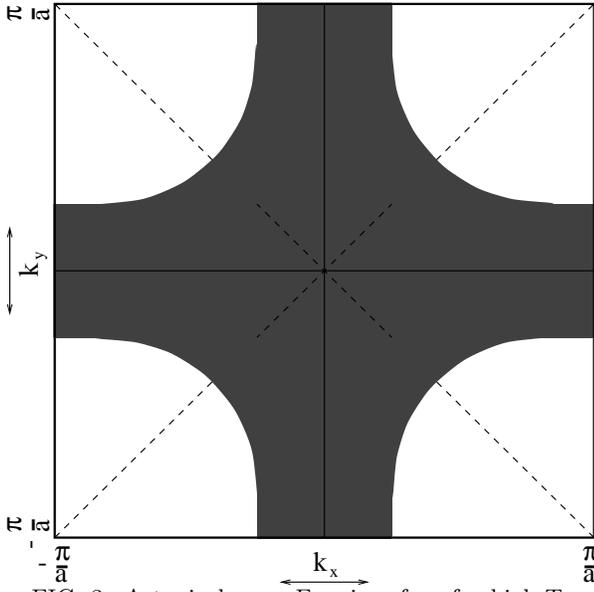,width=8cm,angle=-90}
\caption{A typical open Fermi surface for high-T$_c$ compounds
which can be parametrized
  by a $t-t'$ model. The shaded region denotes the occupied
states with concentration
  $1-x$, where $x$ is the doping concentration. The
superconducting state is gapless
  along the diagonals of the Brillioun zone}
\label{fs}
\end{figure}

Using Eqs. (5) and (4) in Eq.(3) for $H$, and expanding upto quadratic order
in the small quantity ${\bf v}_{s}({\bf r})$ we have
\begin{equation}
H = H_0+H_I+H_{II}  \, ,
\label{eq6}
\end{equation}
where the free Hamiltonian
\begin{equation}
H_0  =
\sum_{{\bf k},\sigma}\xi_{\bf k} c_{{\bf k} \sigma}^\dagger
c_{{\bf k}\sigma}
  +\sum_{\bf k} [\Delta_{{\bf k}} c_{{\bf k}\uparrow}^\dagger
c_{-{\bf k}\downarrow}^\dagger
 +h.c.]  \, ,
\label{eq7}
\end{equation}
with $\xi_{{\bf k}} = -2t[ \cos (k_xa)+\cos (k_y a)] + 4t'\cos
(k_xa)\cos (k_ya) -\mu$ and
$\Delta_{{\bf k}} = (\Delta_0 /2)[ \cos {(k_x a)}-\cos {(k_y
a)}]$, $\Delta_0$
being the maximum quasiparticle excitation gap. 
Here ${\bf k}$ lies in the first atomic Briliouin zone, i.e., 
$-\frac{\pi}{a}\leq (k_x,\, k_y) \leq \frac{\pi}{a}$.
A typical structure of the Fermi surface is shown in Fig.~\ref{fs}. Gapless
quasiparticle excitations exist along $k_x=\pm k_y$ directions as
noted in the
figure.
The interaction term (first order in $v_s$) can now be expressed
as
\begin{eqnarray}
H_I &=& 2(at/\hbar)\sum_{{\bf k}}\sum_{{\bf G}>0} [c_{{\bf
k}\sigma}^\dagger c_{{\bf k}
+{\bf G}\sigma}- c_{{\bf k}+{\bf G} \sigma}^\dagger c_{{\bf
k}\sigma}]\, 
[m\,v_s^x({\bf G})\sin (k_xa) + m\, v_s^y ({\bf G}) \sin
(k_ya) ] \nonumber \\
 & & -4(at' /\hbar) \sum_{{\bf k}}\sum_{{\bf G}>0} [c_{{\bf
k}\sigma}^\dagger c_{{\bf k}
+{\bf G}\sigma}- c_{{\bf k}+{\bf G} \sigma}^\dagger c_{{\bf
k}\sigma}]\, 
\left[ m\,v_s^x({\bf G})
\sin (k_xa) \,\cos (k_y a) \right. \nonumber  \\
& & \left. +m\,v_s^y({\bf G})
\sin (k_ya) \,\cos (k_x a) \right] \nonumber  \\  
&\equiv& \sum_{{\bf k\sigma}}\,\sum_{G>0} V_{{\bf k},{\bf G}}\,[c_{{\bf
k}\sigma}^\dagger
c_{{\bf k}+{\bf G}\sigma}- c_{{\bf k}+{\bf G} \sigma}^\dagger
c_{{\bf k}\sigma}]\, 
\label{eq8}
\end{eqnarray}
Here $V_{{\bf k},{\bf G}}$ is purely imaginary.
The term $H_{II}$ is quadratic in $v_s$ and
contributes to the free energy as a diamagnetic term. It is given by 
\begin{equation}
H_{II} = 2N_s^0 \, (t-t')\, (a^2/\hbar^2) \sum_{{\bf G}}
m\,v_s^\alpha ({\bf G})\, 
m v_s^\alpha (-{\bf G})
\label{eq9}
\end{equation} 
with $N_s^0$ being the number of superfluid carriers, and
$\alpha$ 
refers to cartesian variables $x$ and $y$. Paramekanti {\it et
al.}
\cite{prrm}
have recently shown that the quantum
phase fluctuation of the order parameter reduces the superfluid
density considerably.
We thus reexpress the term $H_{II}$ phenomenologically in terms
of the measured $\lambda$ as
\begin{equation}
H_{II} = \frac{dA}{2\lambda^2}\left( \frac{c^2}{4\pi e^2}\right)
\sum_{{\bf G}} m\,v_s^\alpha ({\bf G})\, 
m v_s^\alpha (-{\bf G}) \, ,
\label{eq10}
\end{equation} 
where $d$ is the mean interlayer separation of weakly coupled
superconducting
layers and $A$ is the area of the system.

\subsection{Free Energy}

We now calculate the free energy as a power series in ${\bf v}_{s}({\bf r})$
or equivalently the vortex density.  The diamagnetic or Ginzburg-Landau term,
of first order in $H_{II}$, is the largest, and the structure sensitive part
of it is known to be minimized for a triangular lattice (1).  The energy does
not depend on its orientation with respect to the crystal lattice.  We are
interested here additionally in the paramagnetic term, of second order in
$H_I$.  Including this, and the magnetic field energy contribution, the free
energy to second order in vortex density is given (per unit length along the
$c$ axis) by \begin{equation} 
\Delta\Omega = \frac{1}{2Ad\hbar^2}\sum_{{\bf G}}
\,mv_{s}^{\alpha}({\bf G})\left[ \chi^d \delta_{\alpha\beta}
  -\chi^{{\rm p}}_{\alpha\beta}({\bf G})\right]
mv_{s}^{\beta}(-{\bf G})
+\frac{1}{8\pi A}\sum_{{\bf G}}B_{{\bf G}}\,B_{-{\bf G}}
\, , \label{eq11} \end{equation} 
where ${\bf G}$ is the reciprocal vector
of the vortex lattice. The individual vortex energy is not included here, as
it is not relevant for the question of vortex lattice structure. 
$\chi^d = (c^2\hbar^2d/4\pi e^2\lambda^2)$ is the diamagnetic term arising from the term
$H_{II}$ (Eq.\ref{eq10}) to first order, and $\chi^{{\rm p}}_{\alpha\beta} ({\bf G})$ is
the paramagnetic current susceptibility 
due to second order contribution from $H_I$. Higher order contributions are
neglected since the expansion parameter is $(n_{v}/n)$ where $n_{v}$ is the
vortex density and $n$ is  the electron density.  This ratio is obviously much
smaller than one.

The paramagnetic susceptibility $\chi^{{\rm p}}_{\alpha\beta} ({\bf q})$ is expressed as
\begin{equation}
\chi^{{\rm p}}_{\alpha\beta} ({\bf q}) = \frac{1}{(2\pi)^2}
\int_{-\pi /a}^{\pi /a}dk_x\int_{-\pi /a}^{\pi /a}dk_y
\gamma_{\alpha\beta} ({\bf k})
\Pi ({\bf k},{\bf q}) \, . 
\label{eq12} 
\end{equation}
The current operator dependent terms
$\gamma_{\alpha\beta}({\bf k})$ are explicitly given as
\begin{mathletters}
\begin{eqnarray} \gamma_{xx} ({\bf k}) &=& \left[ 2a \sin (k_xa) (t-2t' \cos
(k_y a))\right]^2\, ,
\label{eq13a}
\\ 
\gamma_{yy} ({\bf k}) &=& \left[ 2a \sin (k_ya) (t-2t' \cos (k_x
a))\right]^2\, , 
\label{eq13b}
\\
\gamma_{xy} ({\bf k}) &=& \left[2a \sin (k_xa) (t-2t' \cos (k_y
a))\right] 
\left[ 2a \sin (k_ya) (t-2t' \cos (k_x a))\right] \, ,
\label{eq13c} \\
& = & \gamma_{yx} ({\bf k}) \, .
\label{eq13d}
\end{eqnarray}
\end{mathletters}
The zero frequency susceptibility of wave vector ${\bf q}$ for quasiparticle
quasihole of momentum ${\bf k}$ is $\Pi({\bf k},{\bf q})$ and has the form
\begin{equation} 
\Pi ({\bf k},{\bf q}) =  \frac{1}{E_{{\bf k}}+E_{{\bf k}+{\bf
q}}}
\left[ 1 - \frac{\xi_{{\bf k}}\xi_{{\bf k}+{\bf q}}+\Delta_{{\bf
k}}
\Delta_{{\bf k}+{\bf q}}}{E_{{\bf k}}E_{{\bf k}+{\bf q}}} \right]
\, , 
\label{eq14}
\end{equation}
with the quasiparticle energy $E_{\bf k} = \sqrt{\xi_{{\bf k}}^2
+\Delta_{{\bf k}}^2}$.

It is expected that the above susceptibilities are anisotropic
due to the nonlocal
nature of $\Delta_{ij}$, reflected in the ${\bf k}$ dependence of
$\Delta_{{\bf k}}$. Though anisotropic, they possess certain symmetries:
$\chi^{{\rm p}}_{\alpha\alpha}(q_x,-q_y) = \chi^{{\rm p}}_{\alpha\alpha}(q_x,q_y)=
\chi^{{\rm p}}_{\alpha\alpha}(-q_x,q_y)$; $\chi^{{\rm p}}_{xy}(q_x,-q_y)
=-\chi^{{\rm p}}_{xy}(q_x,q_y) = \chi^{{\rm p}}_{xy}(-q_x,q_y)$, and
$\chi^{{\rm p}}_{xx}(q_x,q_y)
=\chi^{{\rm p}}_{yy}(q_y,q_x)$.
These symmetries suggest that the susceptibilities are functions
of $\vert q_x\vert$,
$\vert q_y\vert$ and sign($q_xq_y$) only.
A naive perturbative expansion of $\chi^{{\rm p}}_{\alpha\beta} ({\bf q})$
in powers
of $q$ fails since the
coefficient of quadratic term in $q$ is divergent, due to the 
vanishing of 
$\Delta_{{\bf k}}$ on the Fermi surface at $k_x =\pm k_y$ points. 
We however proceed to evaluate these analytically as follows.

We write
\begin{equation}
\chi^{{\rm p}}_{\alpha\beta}(q_x,q_y) = \sum_{j=1}^4 
   \chi^{{\rm p},j}_{\alpha\beta}(q_x,q_y) \, ,
\label{eq15}
\end{equation}
where $\chi^{{\rm p},j}_{\alpha\beta}$ is the contripution of $j$-th
quadrant $(j=$1--4) of $k$-space. For instance,
\begin{equation}
{\chi}^{{\rm p},1}_{\alpha\beta} (q_x,q_y) = \frac{1}{(2\pi)^2}
\int_{0}^{\pi /a}dk_x\int_{0}^{\pi /a}dk_y \gamma_{\alpha\beta} ({\bf k})
\Pi ({\bf k},{\bf q}) 
\label{eq16}
\end{equation}
is the contribution due to 1st quadrant.
We present the calculation of ${\chi}^{{\rm p},1}_{\alpha\beta} (q_x,q_y)$ 
below in detail.

\begin{figure}[tb]
\psfig{figure=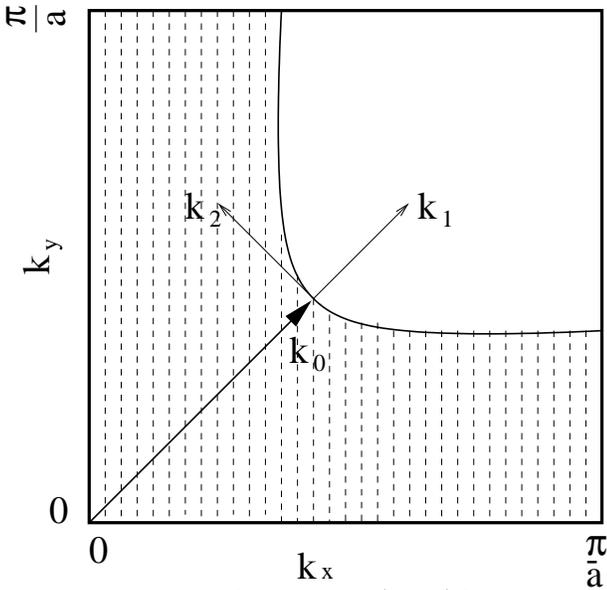,width=8cm,angle=-90}
\caption{New coordinate system $(k_1,k_2)$ in the first quadrant
of
 the atomic BZ is shown. Its origin is at the `nodal' point at
which
 the superconducting gap vanishes on the Fermi surface. The
length
 of the nodal vector is $k_0$.}
\label{node}
\end{figure}

In terms of an alternative coordinate system $(k_1, k_2)$ whose
origin is at the 
nodal point on the Fermi surface as shown in Fig.~\ref{node}, 
the old coordinates in the first
quadrant are expressed as $k_x = \frac{1}{\sqrt{2}}(k_0+k_1-k_2)$
and 
$k_y = \frac{1}{\sqrt{2}}(k_0+k_1+k_2)$, where $k_0$ is defined as
$\mu = -4t\cos (k_0a/\sqrt{2}) + 4t' \cos ^2(k_0a/\sqrt{2})$.
We use
$\xi_{{\bf k}}
\approx \hbar v_F k_1$ and $\Delta_{{\bf k}} \approx \hbar
v_\Delta k_2$ in linear
form, where the Fermi velocity $v_F =
(\frac{4a}{\hbar\sqrt{2}})[t-2t'\cos \frac{k_0a}{
\sqrt{2}}]\sin \frac{k_0a}{\sqrt{2}}$ and $v_\Delta =
(\frac{a}{\hbar\sqrt{2}})
\Delta_0\sin \frac{k_0a}{\sqrt{2}}$ 
is the velocity of quasiparticles along the $k_2$ 
direction. Since in $d$-wave superconductors $v_F \gg v_\Delta$,
the phase space of $k_2$ effectively is
much larger than that of $k_1$ for a given value of quasiparticle
energy. We observe that 
this is the cause of strong anisotropy in the diagonal
susceptibilities as we see below. 
If $\phi$ be the angle of a ${\bf k}$-vector with $k_1$ axis in this 
new coordinate system, $\xi_{{\bf k}} \simeq E_{{\bf k}} \cos \phi$
and $\Delta_{{\bf k}} \simeq E_{{\bf k}} \sin \phi$. By Taylor expansion
in $q$ we find $\xi_{{\bf k}}\xi_{{\bf k+q}}+\Delta_{{\bf k}}\Delta_{{\bf k+q}}
\simeq E_{{\bf k}}(E_{{\bf k}}+\alpha_{{\bf q}})$ and
$E_{{\bf k}}E_{{\bf k+q}} \simeq \vert E^2_{{\bf k}}+E_{{\bf k}}\alpha_{{\bf q}}
+\beta^2_{{\bf q}} \vert $, where $\alpha_{{\bf q}} = \hbar (q_1v_F\cos \phi
+q_2v_\Delta \sin \phi)$ and $\beta^2_{{\bf q}} = (\hbar^2/2)(q_1v_F\sin \phi
-q_2v_\Delta \cos \phi)^2$ with 
$q_{1,2} = (q_y\pm q_x)/\sqrt{2}$ respectively.
Since $\alpha_{{\bf q}}$ is negative for some region of $\phi$,
the quantity $(E^2_{{\bf k}}+E_{{\bf k}}\alpha_{{\bf q}}+\beta^2_{{\bf q}})$
may be negative as well as positive which we refer below as the region I and II 
respectively. It is negative in the regime $E_1^0 < E_{{\bf k}}
< E_2^0$, where $E_1^0 \approx (-\beta^2_{{\bf q}}/\alpha_{{\bf q}})$ 
and $E_2^0 \approx (-\alpha_{{\bf q}}+\beta^2_{{\bf q}}/\alpha_{{\bf q}})$.
Expanding $\gamma_{\alpha\beta}({\bf k})$ upto linear in $E_{{\bf k}}$, we perform
the integrals over $E_{{\bf k}}$ in for both regions I and II separately to
obtain
\begin{mathletters}
\begin{eqnarray}
\chi^{{\rm p},1}_{xx} (q_x,q_y) \approx 
     \frac{a^2}{\pi^2\hbar^2 v_Fv_\Delta} & & \left[ \int_I d\phi \,\left\{
 {\cal R}_1 \left( -\alpha_{{\bf q}}+2(\beta^2_{{\bf q}}/\alpha_{{\bf q}})
\ln |\alpha_{{\bf q}}/\Delta_0| \right)
\right. \right.
\nonumber \\
& & \left. \left. +\frac{2}{3} ({\cal D}^+_\phi \,{\cal R}_2 
+ {\cal D}^-_\phi \,{\cal R}_3)
\left( \alpha^2_{{\bf q}}-3\beta^2_{{\bf q}}
-3\beta^2_{{\bf q}}\ln |\alpha_{{\bf q}}/\Delta_0| \right) \right\}
\right.
\nonumber \\
 & & \, \left. + \int_{II} d\phi \, 
 \left\{ 2\ln (2)\, (\beta^2_{{\bf q}}/\alpha_{{\bf q}}) {\cal R}_1
-2\beta^2_{{\bf q}} ({\cal D}^+_\phi \,{\cal R}_2 + {\cal D}^-_\phi \,{\cal R}_3)
\ln (\alpha_{{\bf q}}/\Delta_0) 
  \right\} \right] \, ,
\label{eq17a} \\
\chi^{{\rm p},1}_{yy} (q_x,q_y) \approx
     \frac{a^2}{\pi^2\hbar^2 v_Fv_\Delta} & & \left[ \int_I d\phi \,\left\{
 {\cal R}_1 \left( -\alpha_{{\bf q}}+2(\beta^2_{{\bf q}}/\alpha_{{\bf q}})
\ln |\alpha_{{\bf q}}/\Delta_0| \right)
\right. \right. 
\nonumber \\
& & \left. \left.
 +\frac{2}{3} ({\cal D}^+_\phi \,{\cal R}_3 + {\cal D}^-_\phi \,{\cal R}_2)
\left( \alpha^2_{{\bf q}}-3\beta^2_{{\bf q}}
-3\beta^2_{{\bf q}}\ln |\alpha_{{\bf q}}/\Delta_0| \right) \right\}
\right.
\nonumber \\
 & & \, \left. + \int_{II} d\phi \,
 \left\{ 2\ln (2)\, (\beta^2_{{\bf q}}/\alpha_{{\bf q}}) {\cal R}_1
-2\beta^2_{{\bf q}} ({\cal D}^+_\phi \,{\cal R}_3 + {\cal D}^-_\phi \,{\cal R}_2)
\ln (\alpha_{{\bf q}}/\Delta_0)
  \right\} \right] \, ,
\label{eq17b} \\
\chi^{{\rm p},1}_{xy} (q_x,q_y) \approx
     \frac{a^2}{\pi^2\hbar^2 v_Fv_\Delta} & & \left[ \int_I d\phi \,\left\{
 {\cal R}_1 \left( -\alpha_{{\bf q}}+2(\beta^2_{{\bf q}}/\alpha_{{\bf q}})
\ln |\alpha_{{\bf q}}/\Delta_0| \right)
\right. \right.
\nonumber \\
& & \left. \left.
 +\frac{2}{3} ({\cal R}_2 + {\cal R}_3)\cos \phi\,
\left( \alpha^2_{{\bf q}}-3\beta^2_{{\bf q}}
-3\beta^2_{{\bf q}}\ln |\alpha_{{\bf q}}/\Delta_0| \right) \right\}
\right.
\nonumber \\
 & & \, \left. + \int_{II} d\phi \,
 \left\{ 2\ln (2)\, (\beta^2_{{\bf q}}/\alpha_{{\bf q}}) {\cal R}_1
-2\beta^2_{{\bf q}} ({\cal R}_2 + {\cal R}_3) \cos \phi \,
\ln (\alpha_{{\bf q}}/\Delta_0)
  \right\} \right] \, ,
\label{eq17c}
\end{eqnarray}
\end{mathletters}
where ${\cal D}^\pm_\phi = \cos\phi \pm (v_F/v_\Delta)\sin \phi$,
\begin{mathletters}
\begin{eqnarray}
{\cal R}_1 &=& (t^2+\mu t' )\sin ^2(k_0a/\sqrt{2}) \, , 
\label{eq18a} \\ 
{\cal R}_2 &=& 2\sqrt{2}(\frac{a}{\hbar v_F})t'(t-2t'\cos (k_0a/\sqrt{2}))
                 \sin^3(k_0a/\sqrt{2}) \, , 
\label{eq18b} \\ 
{\cal R}_3 &=& \sqrt{2}(\frac{a}{\hbar v_F})(t^2+\mu t')\sin (k_0a/\sqrt{2})
                 \cos (k_0a/\sqrt{2}) \, ,
\label{eq18c}
\end{eqnarray}
\end{mathletters}
and $\int_I$ and $\int_{II}$ represent the integrals over $\phi$ 
$(0\leq \phi \leq 2\pi)$ for the regime of $\phi$ in which
$\alpha_{{\bf q}} < 0$ and $>0$ respectively.
We, similarly, calculate $\chi^{{\rm p},j}_{\alpha\beta}$ for three other quadrants.
We observe that $\chi^{{\rm p},1}_{xx}$ and 
$\chi^{{\rm p},1}_{yy}$ differ substantially through the last terms
within both the integrals $\int_{I}$ and $\int_{II}$ in
their expressions (\ref{eq17a}) and (\ref{eq17b}) since $v_F \gg
v_\Delta$. 
These lead to an
anisotropic diagonal susceptibility. We note that these terms
also the terms involving $\int_{II}$ arise due to keeping the 
linear dependences of $k_{1}$ and $k_{2}$ in
$\gamma_{\alpha\beta}({\bf k})$. 
However for an approximation $k_{x}=k_{y}=k_{0}/\sqrt{2}$ 
 in $\gamma_{\alpha\beta} ({\bf k})$,
$\chi^{{\rm p}}_{xx}=\chi^{{\rm p}}_{yy}$ as obtained by Franz
et al \cite{franz}.

Since the angular integrals in the expressions for
$\chi^{{\rm p}}_{\alpha\beta}$
cannot be performed analytically, we numerically integrate these
to obtain
$\chi^{{\rm p}}_{\alpha\beta}$ in the next section. We shall then compare
these semianalytically 
obtained $\chi^{{\rm p}}_{\alpha\beta}$ with those completely numerically
obtained through Eqs.(\ref{eq12})--(\ref{eq14}).

We now turn to obtain the equation for the vector potential in
the
gauge ${\bf G}\cdot {\bf A}_{{\bf G}}=0$ (from Eq.~(\ref{eq11})). 
By minimizing the energy  with respect to as ${\bf A}_{{\bf G}}$, we have
\begin{equation}
(A_{{\bf G}})_\alpha = \frac{4\pi e}{\hbar
cG^2d}[\chi^d \delta_{\alpha\beta} -\chi^{{\rm p}}_{\alpha\beta}({\bf G})]
\left[ \frac{1}{2}(\nabla\theta)_{-{\bf G}}-\frac{e}{\hbar
c}A_{-{\bf G}}\right]_\beta \, .
\label{eq19}
\end{equation}
We thus obtain
\begin{equation}
\left[ \frac{\hbar}{2}(\nabla\theta)_{-{\bf G}}\right]_\alpha
 = \left[ G^2 Q^{-1}_{\beta\alpha} +\delta_{\beta\alpha}\right]
\left( \frac{e}{c}A_{-{\bf G}} \right)_\beta \, 
\label{eq20}
\end{equation}
where
\begin{equation}
Q_{\alpha\beta} ({\bf q}) = \frac{1}{\lambda^2}\delta_{\alpha\beta} -\left( \frac{4\pi
e^2}{c^2\hbar^2d}\right) \chi^{{\rm p}}_{\alpha\beta} ({\bf q})
\label{eq21}
\end{equation}
Using Eq.~(\ref{eq20}) in Eq.~(\ref{eq11}), we get
\begin{equation}
\Delta\Omega = \frac{1}{8\pi A} \sum_{{\bf G}} B_{{\bf G}} 
\left[ 1+
\frac{G_\alpha Q_{\alpha\beta}G_\beta}{\mbox{Det}
\, Q } \right] 
B_{-{\bf G}}\, .
\label{eq22}
\end{equation}

We now express $B_{{\bf G}}$ in terms of $N_{v}, \Phi_{0}, Q_{\alpha\beta}$ and
${\bf G}$.  For a vortex lattice, 
\begin{equation}
({\bf \nabla}\theta )_{{\bf G}} = 2i\pi N_v \frac{{\bf G}\times
\hat{e}_z}
{G^2} \, ,
\label{eq23}
\end{equation}
where $N_v$ is the total number of vortices. 
We then obtain from Eqs.~(\ref{eq19}) and (\ref{eq20})
\begin{equation}
B_{{\bf G}} = N_v\Phi_0\left[ 
\frac{\mbox{Det}\, Q+ Q_{\alpha\beta}\epsilon^{\alpha
\gamma}\epsilon^{\beta\delta} G_\gamma G_\delta}{G^4 +G^2
Q_{\alpha\beta}
\delta_{\alpha\beta} +\mbox{Det}\, Q} \right] \, ,
\label{eq24}
\end{equation}
with $\epsilon^{12}=1=-\epsilon^{21}$,
$\epsilon^{11}=0=\epsilon^{22}$, and $\Phi_0 = hc/2e$ is the
quantum of flux.
Therefore, the free energy for a vortex lattice per unit volume
becomes
\begin{equation}
{\cal F} = \frac{1}{8\pi}(\Phi_0 n_v)^2\sum_{{\bf G}}
\left[ \frac{\mbox{Det}\, Q+ Q_{\alpha\beta}\epsilon^{\alpha
\gamma}\epsilon^{\beta\delta} G_\gamma G_\delta}{G^4 +G^2
Q_{\alpha\beta}
\delta_{\alpha\beta} +\mbox{Det}\, Q} \right]^2
\left[ 1+
\frac{G_\alpha Q_{\alpha\beta}G_\beta}{\mbox{Det}
\, Q } \right] 
 \,.
\label{eq25}
\end{equation}
This has an approximate but much simpler form as
\begin{equation}
{\cal F} \simeq \frac{1}{8\pi}(\Phi_0 n_v)^2 \sum_{{\bf G}}
\frac{G_x^2 Q_{yy}+G_y^2 Q_{xx}-G_xG_y (Q_{xy}+Q_{yx})}{G^4} \, ,
\label{eq26}
\end{equation}
which is essentially important for determining the ground state
structure of the vortex lattice.
This form is exact when $\vert \frac{\hbar}{2}(\nabla
\theta)_{{\bf G}}\vert
\gg \vert eA_{{\bf G}}\vert /c$ which is true.
The component $G=0$ will give free energy $\bar{{\cal F}}$
for average magnetic induction $B$.
For determination of vortex lattice structure, one should in
principle minimize Gibbs
free energy ${\cal G} = {\cal F}-BH/4\pi$. Here $H$ is the
applied 
magnetic field.
Beyond $H_{c_1}$, magnetic field penetrates the superconductor
almost fully. Thus $B
\simeq H$, specially so in high $T_c$ superconductors, since $H_{c_1}\ll
H_{c_2}$. 
$B$ does not vary much for 
different vortex lattice structures for a given $H$ as we see in 
our numerical study that the ratio $(H-B)/B\sim 10^{-7}$. 
We, therefore,
minimize $\Delta{\cal F}={\cal F}-\bar{{\cal F}}$ ie. that part of the free
energy which depends on $G$, in Eq.~(\ref{eq26})  for different choices of
nonzero $G$'s corresponding to different structures
and with a cutoff $G\leq \pi/\xi$.

\section{Numerical Study}

 The values of the phenomenological parameters that we have used 
for the numerical computation of $Q_{\alpha\beta}(q_x,q_y)$ and later for the 
free energy are taken from angle resolved photoemission
experiments \cite{norman2,norman1}, penetration depth measurement
\cite{bonn2}
and band structure calculations \cite{dagotto} for
high-T$_c$ compounds. These are as follows: $t=1150$K,
$t'=0.48t$, $\Delta_0=400$K,
$a=3.8\AA$, $d=10\AA$, $\xi=20\AA$, $\lambda = 1600\AA$,
and $\mu = -1.33t$ which corresponds to doping concentration $x\simeq 0.19$.

Using standard Gaussian quadrature, we integrate over $k_x$ and $k_y$ in
Eq.~(\ref{eq12}) to obtain $\chi^{{\rm p}}_{\alpha\beta} ({\bf q})$. In
Fig.~\ref{chiab} we have shown the
dependence of paramagnetic susceptibilities (a) $\chi^{{\rm
p}}_{xx}$, (b) $\chi^{{\rm p}}_{yy}$, and
(c) $\chi^{{\rm p}}_{xy}$ in units of $\chi^d$ for positive $q_x$ at different
positive values of $q_y$. Susceptibilities for negative values of $q_x$ and
$q_y$ can be obtained by using the symmetries discussed in the previous
section. We see that $\chi^{{\rm p}}_{xx} (q_x, q_y) \neq \chi^{{\rm p}}_{yy}(q_x, q_y)$ in general.
This strong anisotropy in the diagonal susceptibilities is due to the strong
${\bf k}$ dependence of nature of $\Delta_{{\bf k}}$ and the ${\bf k}$-dependent
$\gamma_{\alpha\beta}({\bf k})$  (\ref{eq13a})--(\ref{eq13d}). The diagonal
susceptibilities are large compared to off-diagonal one.

We numerically fit, guided by the semianalytical form in 
Eqs.(\ref{eq17a})--(\ref{eq17c}), to obtain the approximate functional form 
of $\chi^{{\rm p}}_{\alpha\beta} (q_x,q_y)$ for $q_xa,q_ya \leq 0.3$ as
\begin{mathletters}
\begin{eqnarray}
\chi^{{\rm p}}_{xx}(q_x,q_y) &=&  \left\{  \begin{array}{lll}
\gamma \big[ 0.31 (\delta\vert q_x\vert a) + 0.14 (\delta
q_xa)^2 -0.35 (\delta q_x a)^2
 \ln \vert \delta q_x a\vert \big] &  {\rm if} & \vert
q_x\vert \geq \vert q_y\vert , \\
 \gamma \Big[ 0.35 (\delta\vert q_y\vert a) - 0.14 (\delta
q_ya)^2 +0.10 (\delta q_y a)^2
 \ln \vert \delta q_y a\vert  &  &   \\
  +(0.10+\frac{0.21}{\delta\vert q_y\vert a})(\delta
q_x a)^2
 +(-0.16+\frac{0.07}{\delta\vert q_y\vert a})(\delta q_x
a)^2\ln \vert
 \delta q_x a\vert \Big]
 &  {\rm if} & \vert q_x \vert \leq \vert q_y \vert,
\end{array} \right.
\label{eq27a} \\
\chi^{{\rm p}}_{yy}(q_x,q_y) &=& \chi^{{\rm p}}_{xx}(q_y,q_x)\, , 
\label{eq27b} \\
\chi^{{\rm p}}_{xy}(q_x,q_y) &=& \gamma \left[
(0.11+\frac{0.02}{q_>})q_< +
(0.15-\frac{0.18}{q_>})q_<^2  
 +(0.07-\frac{0.12}{q_>})q_<^2
\ln q_< \right]
 {\rm sign}(q_xq_y)
\label{eq27c}
\end{eqnarray}
\end{mathletters}
with $q_{>,<}={\rm max, min}(\vert q_x\vert, \vert q_y \vert)
\delta a$, $\gamma =
\frac{\lambda^2}{d}(\frac{4\pi e^2}{c^2\hbar^2})t$ whose numerical value is 1.18,
and the parameter $\delta = t/\Delta_0$.
These phenomenological forms can be explained from the semianalytical expressions
(\ref{eq17a})--(\ref{eq17c}) as follows.
(i) Firstly, why do $\chi^{{\rm p}}_{xx} (q_x,q_y)$ and
$\chi^{{\rm p}}_{yy} (q_x,q_y)$ not depend on the signs of $q_x$ and $q_y$,
and  $\chi^{{\rm p}}_{xy} (q_x,q_y)$ does depend on sign$(q_xq_y)$? 
This is due to the symmetry reason discussed following Eq.(\ref{eq14}).
(ii) Why does $\chi^{{\rm p}}_{xx}(q_x,q_y)$ does mainly depend on whether
$\vert q_x \vert \geq \vert q_y \vert$ or not? This can be understood by the
following exercise. We find a term from Eq.(\ref{eq17a}) as $\vert q_y + q_x
\vert$ assuming $q_x,q_y\geq 0$. The corresponding term will be $\vert
q_y - q_x\vert$ when we consider the contribution from 2nd quadrant. When
we add these two contributions, we see that the sum depends on the greater
of $q_x$ and $q_y$. 
(iii) Following the argument above in (ii), the difference between the two
terms is smaller of $q_x$ and $q_y$. This is the reason why 
$\chi^{{\rm p}}_{xy}(q_x,q_y)$ depends mainly on the smaller of $\vert q_x\vert$
and $\vert q_y\vert$.   
(iv), Since $\chi^{{\rm p},1}_{xx} (q_x,q_y) \neq \chi^{{\rm
p},1}_{yy}(q_x,q_y)$, and $\chi^{{\rm p}}_{xx} (q_x,q_y) = \chi^{{\rm
p}}_{yy}(q_y,q_x)$ for symmetry reasons,
the dependence of $\chi^{{\rm p}}_{xx} (q_x,q_y)$ on $q_x$ and $q_y$ is asymmetric.
(v) The linear, quadratic, and the logarithmic dependences on $q$ follow from
in the expressions (\ref{eq17a})--(\ref{eq17c}).

\begin{figure}[tb]
\psfig{file=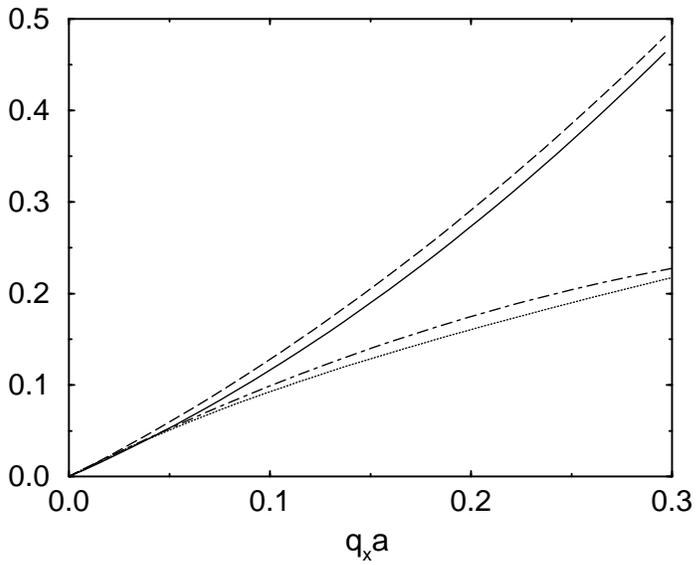,width=12cm,angle=-90}
\caption{Numerically (dashed and dotted lines) and
semianalytically (solid and
dot-dashed lines) obtained $\chi^{{\rm p}}_{xx}(q_x,0)$
  and $\chi^{{\rm p}}_{yy}(q_x,0)$ 
(in units of $\chi_d$)
respectively.}
\label{chi}
\end{figure}

 We next numerically perform angular integrals in equations 
(\ref{eq17a})--(\ref{eq18c}) along with the contributions
from other three quadrants to obtain semianalytical
susceptibilities and 
then compare with the fully numerically obtained susceptibilities.
In Fig.~\ref{chi} we have shown
$\chi^{{\rm p}}_{xx}(q_x,0)$ and $\chi^{{\rm p}}_{yy}(q_x,0)$ evaluated in
the two ways. The linear approximation of energies in the analytical
expressions is a
good approximation for determining linear dependence on $q_x$ as
we see
in Fig.~\ref{chi} that they agree for very low $q_x$. They however differ
for
higher $q_x$ since our analytical expressions  are not consistent
in determining quadratic dependences on $q$ as we have 
neglected higher order $k$ dependences to quasiparticle energy. It is however
clear that $\chi^{{\rm p}}_{xx} \neq \chi^{{\rm p}}_{yy}$ which
is our main result.

We consider a face centered rectangular vortex lattice (as shown
in Fig.~\ref{lat_n})
with area of the
unit cell $\tilde{A}=2\Phi_0/B$, in general. Angle $\theta$ 
determines the sides of the rectangle with a fixed area. 
The sides of the rectangle are
$a_1 = [\tilde{A}\tan \theta ]^{1/2}$ and $a_2=[\tilde{A}/\tan
\theta ]^{1/2}$.
We then readily
obtain reciprocal lattice vectors for a vortex lattice, in
general, 
to be
\begin{equation}
{G}_{mn}(B,\theta) = (n+m)\frac{2\pi}{a_1}\hat{e}_x
+(n-m)\frac{2\pi}
{a_2} \hat{e}_y \, ,
\label{eq28}
\end{equation}
where $n$ and $m$ are integers (both positive and negative)
including zero.
If the vortex lattice makes an angle $\phi$ with the underlying
atomic lattice,
we find
\begin{eqnarray}
{G}_{mn}(B,\theta ,\phi) &=& \hat{e}_x 
\left[ (n+m)\frac{2\pi}{a_1} \cos \phi -(n-m)\frac{2\pi}
{a_2} \sin \phi \right] \nonumber \\
& & +\hat{e}_y \left[ (n+m)\frac{2\pi}{a_1}
\sin \phi +(n-m)\frac{2\pi} {a_2} \cos \phi \right] 
\, .
\label{eq29}
\end{eqnarray}
The lattice is a centered square for $\theta = 45^o$ and triangular
when $\theta = 60^o$. There is symmetry of rotation about $\phi =
45^o$,
since the lattice is considered as centered rectangular. 
We therefore need to determine free energy  
for $45^o\leq \theta < 90^o$ and $0\leq \phi \leq 45^o$.

\begin{figure}[tb]
\psfig{figure=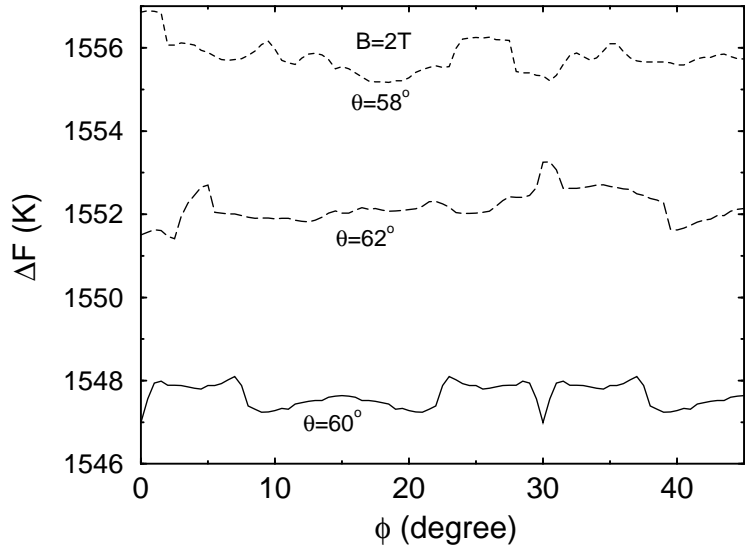,width=12cm,angle=-90}
\caption{Free energy per vortex as a function of $\phi$ for
$\theta=58^o$, $60^o$ and $62^o$
 at 2 Tesla field.}
\label{fe_h2_th}
\end{figure}

 We then numerically compute the free energy per vortex (without
the single vortex energy which does not depend on structure), $\Delta F =
d\Delta {\cal F}/n_v$ using Eqs.(\ref{eq26}) and (\ref{eq29}) 
as functions of the parameters $\theta$, $\phi$, and $B$. The 
reciprocal lattice vector ${\bf G}$ changes with the change of any
one or more of the
parameters. Thus $\Delta F$ differs for different combination of
these parameters $(\phi ,\theta ,H)$.
In Fig.~\ref{fe_h2_th} we have shown the dependence of $\Delta F$ at a low
field $B=2$ Tesla
as a function of $\phi$ for the angles $\theta = 60^o$ and two
neighboring angles 
$\theta = 58^o$ and $62^o$ (on either side of $\theta = 60^o$).
It is clear that $\Delta
F$ is a minimum for the triangular lattice. We notice also that
$\Delta F$ is a minimum for
the triangular lattice when
$\phi =0^o$ and $30^o$ which in fact correspond to the same
lattice configuration. 
We likewise find that in the whole of the low field regime, the
ground state configuration
of the vortex lattice is triangular with one of its arms parallel
to one of the crystal axes.

\begin{figure}[tb]
\psfig{figure=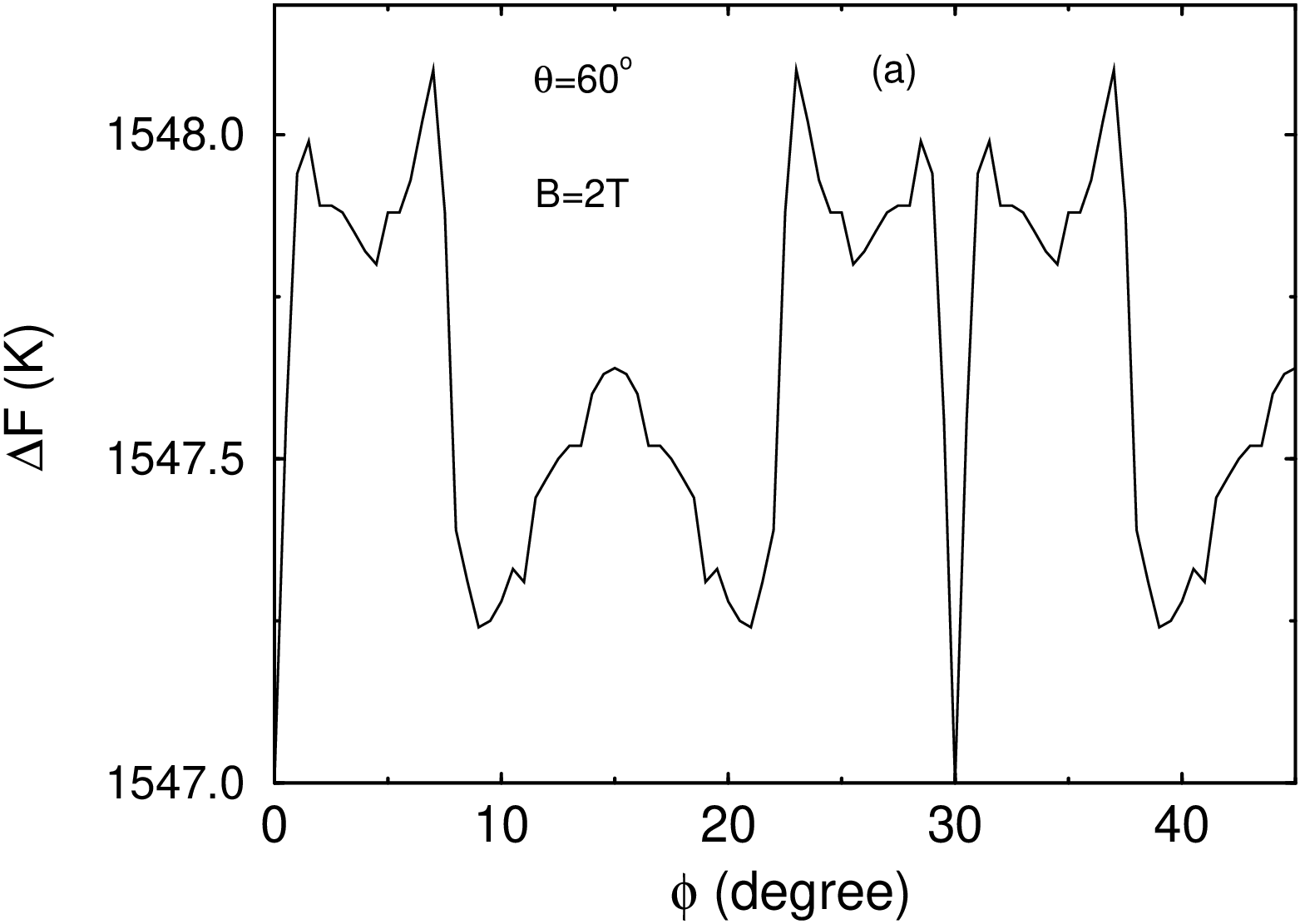,width=12cm,angle=-90}
\psfig{figure=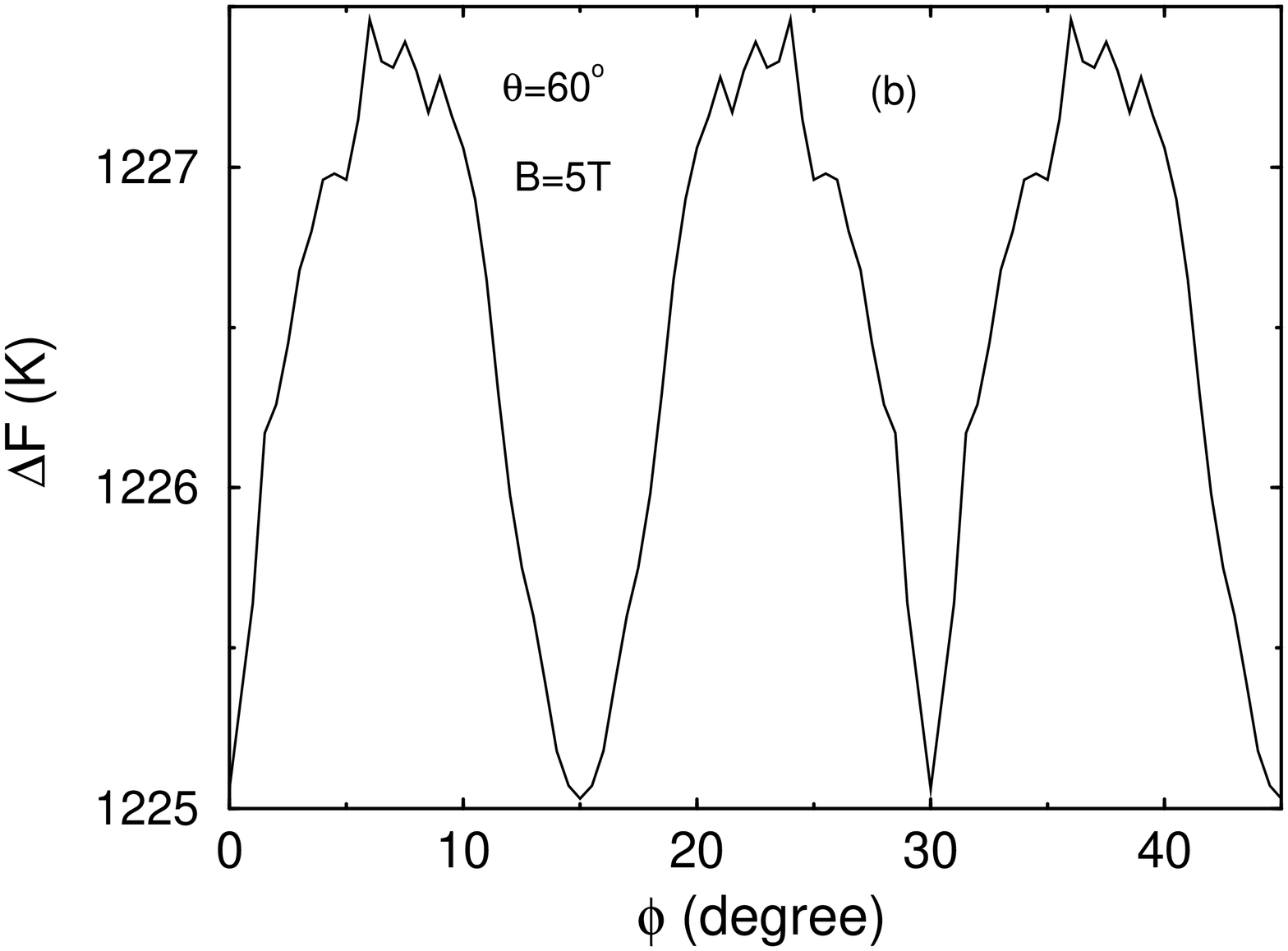,width=12cm,angle=-90}
\psfig{figure=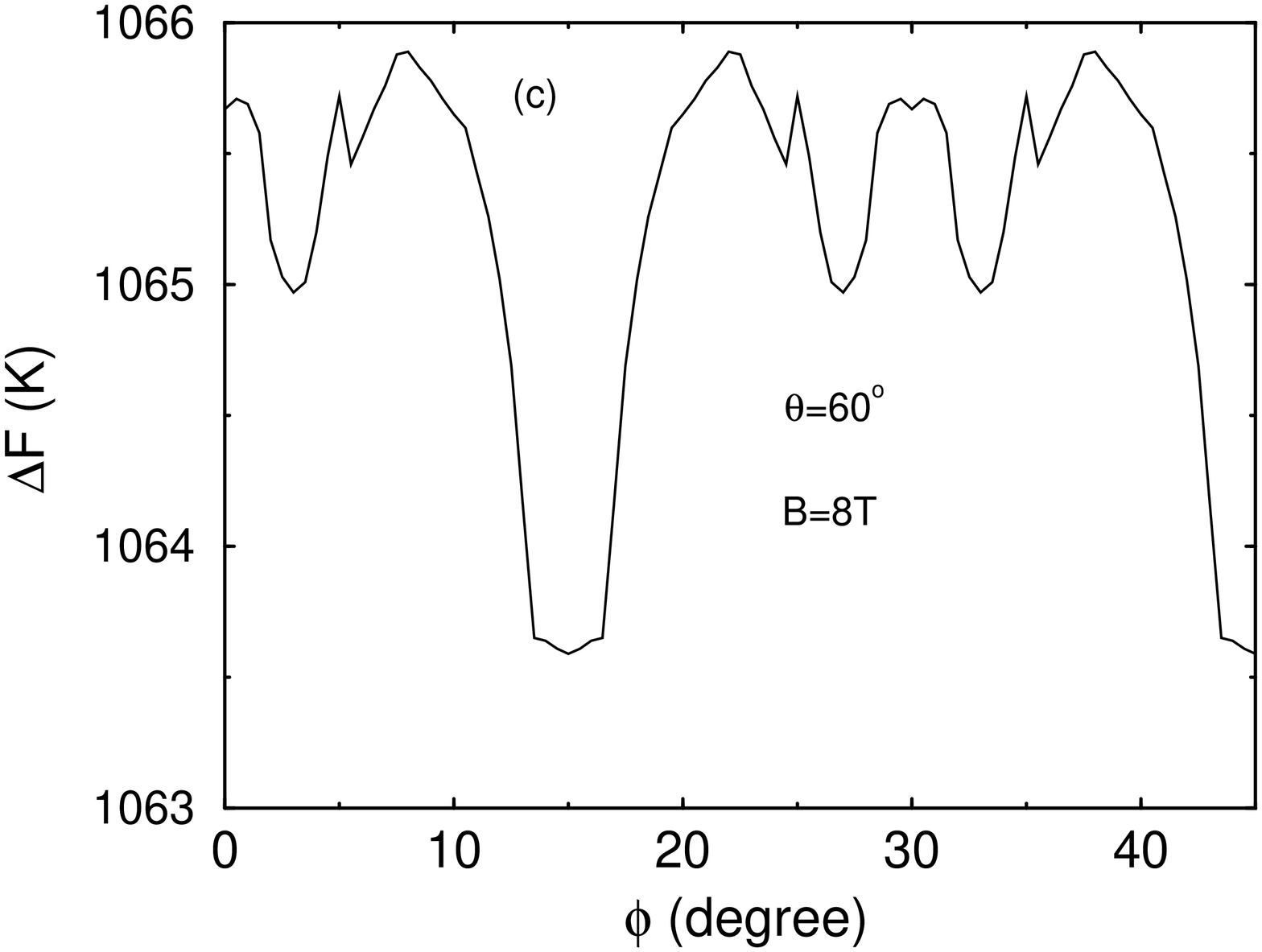,width=12cm,angle=-90}
\caption{Free energy per vortex for triangular lattice structure
($\theta=60^o$) as a function of
 its orientation angles $\phi$ for (a) $B=2$, (b) $B=5$, and (c)
$B=8$ Tesla fields.}
\label{fe_h}
\end{figure}

\begin{figure}[tb]
\psfig{file=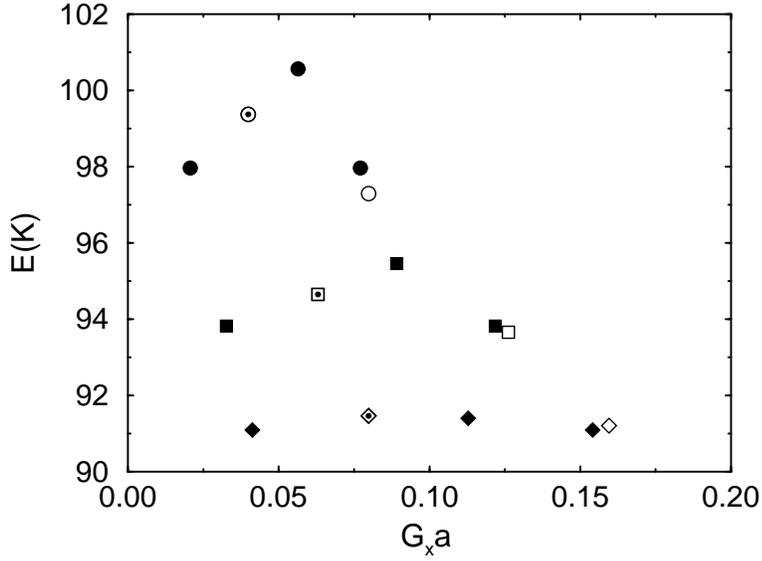,width=12cm,angle=-90}
\caption{The contributions to energy per vortex by the three
lowest ${\bf G}$ vectors
 of equal length
 for positive $G_x$ at three different fields for two different
commensurate orientations
 of triangular lattice. The open(closed) symbols represents
$0^o(45^o)$ orientation of
 vortex lattice with respect to crystalline lattice. The open symbols
with a dot in their centers
 correspond to energy for two different ${\bf G}$ vectors with
same $G_x$. The circles,
 squares and diamonds correspond to 2, 5, and 8 Tesla fields
respectively.}
\label{comp}
\end{figure}

Interestingly, the orientation of the lattice changes
discontinuously as we increase
the magnetic field though the structure continues to be triangular.
In Fig.~\ref{fe_h} we have shown
the dependence of $\Delta F$ on $\phi$ for the triangular lattice
configuration at
three chosen fields 2, 5, and 8 Tesla. At nearly about 5 Tesla
field, $\Delta F$ is minimum
for all $0^o$, $30^o$, $15^o$, and $45^o$ orientations; the
latter two angles correspond to
same lattice configuration, like the former two angles. On the
other hand at the field of 
8 Tesla, $\Delta F$ is minimum for $\phi =15^o$ and $45^o$ only.
The triangular 
vortex lattice changes its orientation discontinuously at about 5
Tesla field. While the
triangular lattice has one of its arms parallel to one of the
crystal axes at lower field,
it aligns to one of the crystal axes by $45^o$ at higher field. 
We understand this discontinuous transition by comparing the
energies contributed to $\Delta F$
by the ${\bf G}$ vectors of the lowest magnitude (since they
contribute most to the free
energy) for these two preferred orientations. 
Considering the symmetries of the susceptibilities, it is
sufficient that we consider only those 
${\bf G}$ vectors which have positive $G_x$. We thus consider 3
${\bf G}$ vectors for each
of these two orientations. These are (a) $(1/2,\pm \sqrt{3}/2)G$
and $(1,0)G$ for $\phi = 0^o$ and
(b) $\frac{1}{2\sqrt{2}}(\sqrt{3}-1, \sqrt{3}+1)G$,
 $\frac{1}{2\sqrt{2}}(\sqrt{3}+1, \sqrt{3}-1)G$, and
$(1/\sqrt{2}, 1/\sqrt{2})G$ for $\phi
=45^o$, where the length of the smallest ${\bf G}$ vector $G =
2\pi (2/\sqrt{3})^{1/2} 
(B/\Phi_0)^{1/2}$. 
In Fig.~\ref{comp}, we have shown the energy {\it E}
contributed by these individual ${\bf G}$ vectors to 
$\Delta F$ for 2, 5, and 8 Tesla fields. 
We find the total energy contributed by above 3 ${\bf G}$
vectors for $\phi = 0^o$ and $45^o$ orientations as (a) 296.05K
and 296.51K for $B=2$
Tesla, (b) 282.96K and 283.11K for $B=5$ Tesla, and (c) 274.15K
and 273.60K for
$B=8$ Tesla respectively. 
Clearly, the triangular lattice makes an orientational transition
at about 5 Tesla field.

\begin{figure}
\psfig{file=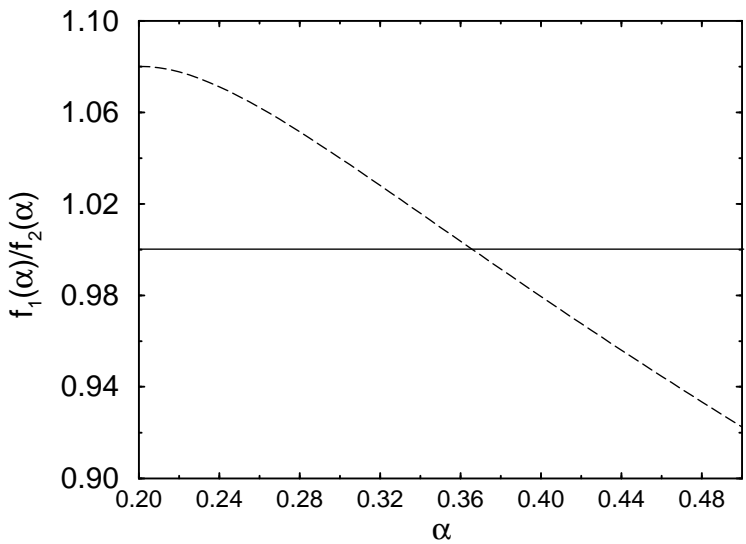,width=12cm,angle=-90}
\caption{The dashed line represents the ratio
$f_1(\alpha)/f_2(\alpha)$ as a
function of $\alpha$. The solid line is a guide to the eyes for
the value 
of the ratio 1.0.}
\label{scale}
\end{figure}

 To understand the field scale 5 Tesla for the above
orientational transition, we
compare the energy contributed by the above
3 ${\bf G}$ vectors for each of the 
preferred orientations. The ratio of these energies can be
expressed as a function of
$\alpha = tGa/\Delta_0$ using equations
(\ref{eq26})--(\ref{eq29}). 
This is given by
\begin{equation}
\frac{E_1}{E_2} = \frac{3-f_1(\alpha)}{3-f_2(\alpha)}
\label{eq30}
\end{equation} 
where $E_1 (E_2)$ is the energy contributed by the above
corresponding 3 ${\bf G}$
vectors of triangular lattice with $0^o (45^o)$ orientation. In
Fig.~\ref{scale}, we have
shown the ratio $f_1(\alpha)/f_2(\alpha)$ as a function of
$\alpha$. The orientational
transition takes place when the ratio is unity. This corresponds
to $\alpha_c =
tG_ca/\Delta_0 \simeq 0.37$. Therefore the critical field at
which the transition
take place, $B_1 \simeq
(0.37/2\sqrt{2}\pi)^2\sqrt{3}(\Delta_0/at)^2\Phi_0 
\simeq 5.2T$.

\begin{figure}
\psfig{file=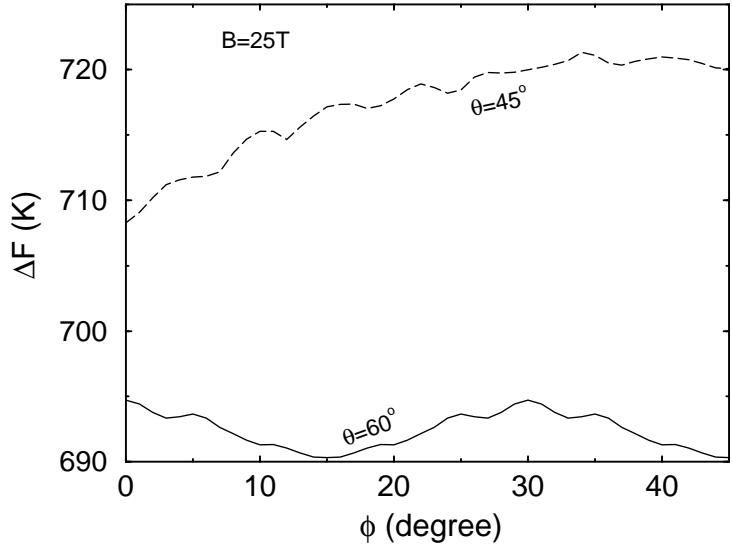,width=12cm,angle=-90}
\caption{Free energy per vortex as a function of $\phi$ for
 $\theta = 60^o$ and $45^o$ at a high field $B=25$ Tesla.}
\label{fe_h25}
\end{figure}

\begin{figure}
\psfig{file=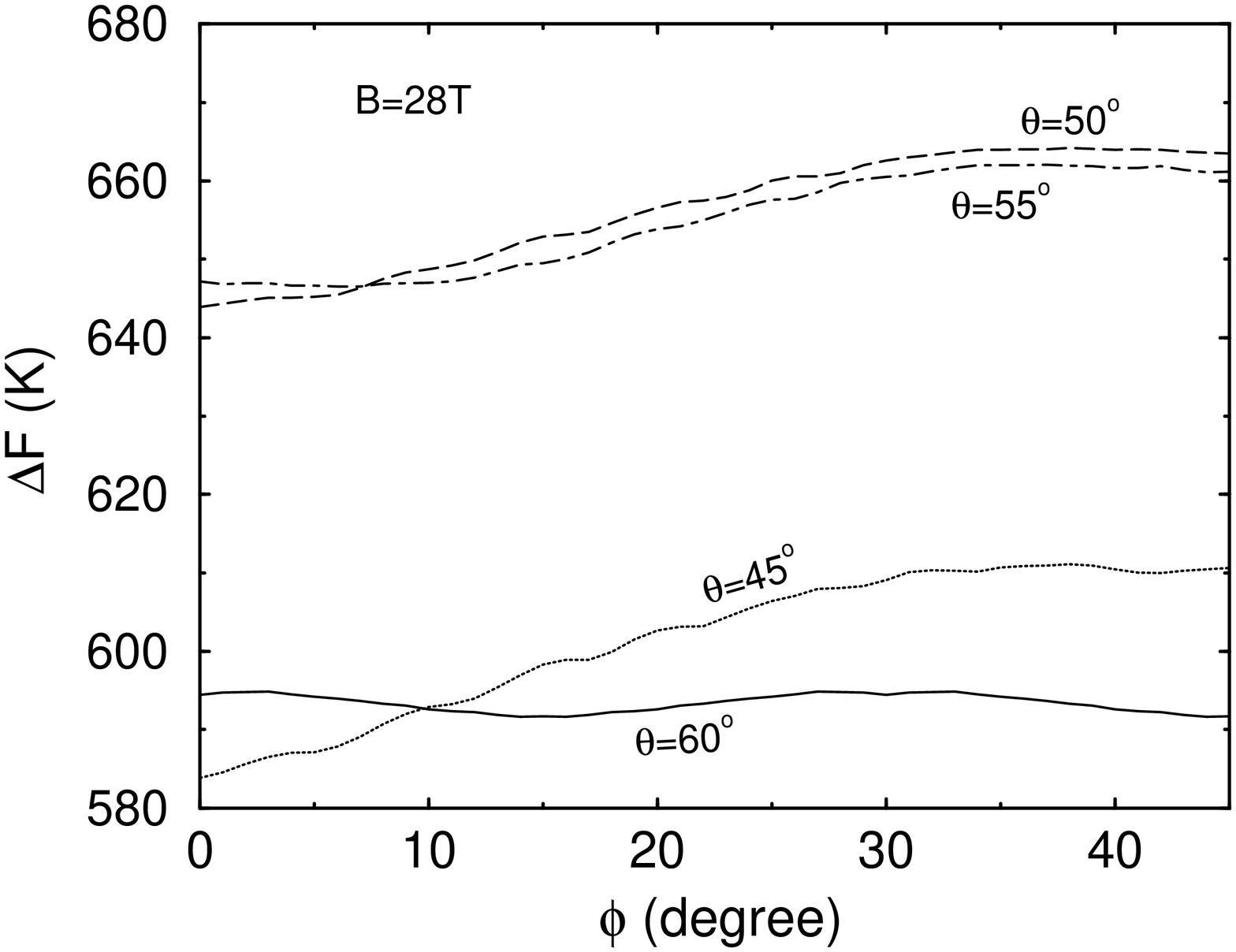,width=12cm,angle=-90}
\caption{Free energy per vortex as a function of $\phi$ for
 $\theta = 45^o$, $50^o$, $55^o$, and $60^o$ at $B=28T$ field.
Clearly
 $\Delta F$ is minimum for the structure corresponding th $\theta
= 45^o$
 and $\phi = 0^o$.}
\label{fe_h28}
\end{figure}

 The structure of the vortex lattice remains triangular
with $45^o$ orientation to 
the crystal lattice as we have shown $\Delta F$ 
in Fig.~\ref{fe_h25} for a field as high as
$25$ Tesla.  However, it makes yet another discontinuous structural transition
to a centered square lattice with
its axes parallel to the crystal axes at yet another critical
field
$B_2$ whose value is about 28 tesla. Figure \ref{fe_h28} shows that 
$\Delta F$ is minimum for $\theta = 45^o$ and $\phi = 0^o$ at $B
= 28$ Tesla.
The overall phase diagram for the ground state of the vortex
lattice structure
at $T=0$ is shown in figure \ref{phase_n}.

\section{Concluding Remarks}

We conclude by briefly discussing a number of questions such as the nature of
the approximations used, the effect of nonzero temperature, consequences of
the transitions, their observability, the reason why our conclusions differ
from those found earlier, and the structural transitions experimentally
observed.

We have calculated the ground state energy assuming effectively that the
interaction between two vortices is unaffected by the presence of other
vortices.  This is obviously a low vortex density approximation which seems
quite reasonable since the dimensionless ratio $(n_{v}/n)$ is about 1/2500 
for a field of a Tesla.  However,
we have not calculated the higher order corrections which while nominally of
higher order in $(n_{v}/n)$ might have large or even divergent coefficients. 
Since the vortex vortex interaction depends on quasiparticle-quasihole
susceptibility, a change in their spectrum due to the supercurrent (the Volovik
effect \cite{volovik} ) could have serious consequences.  Here, we would like to make
two points.  Firstly, a calculation by Amin, Affleck and Franz \cite{amin}, using a
semiclassical approximation to include the nonlinear effect of the magnetic
field {\em a la} Volovik, finds that this has little effect on the structural
transformations calculated by them.  Secondly, {\em all} the recent fully
quantum mechanical calculations \cite{tesanovic,halperin} of the density of Dirac quasiparticle
states in a vortex lattice in the London limit find that for Bravais lattices,
the density of states vanishes linearly with energy as in the absence of a
magnetic field; only the slope changes.  A general argument for this has been
presented by Vishwanath \cite{vishwanath}.  For these two reasons, we believe
that our low density approximation is reliable.

In the London approximation, the vortex cores are treated as $\delta$
functions.  In reality, they have a width of order the coherence length.  We
believe that the consequences of this approximation, at least for the low
field structural transition, are small.  The reason is that the structure
sensitive part of the energy arises from the difference in contribution of the
smallest reciprocal lattice vectors (Section III).  For these, $G\xi
\sim (1/10)$ at typical magnetic fields, so that the phenomenological
assumption of a Gaussian vortex core with width $\sim \xi$ will make a
negligible difference to the structure sensitive part.

We have calculated only the ground state energy of the vortex lattice in this
paper. At any nonzero temperature, there are obviously entropic contributions
which could change the magnetic field at which the structural transition
occurs, as a function of temperature.  Here, we note that since both the
structures (below and above 5T) are identical (triangular), and the structure
difference sensitive part of the energy is a tiny fraction $(<(1/1000)^2)$ of
the vortex interaction energy, the elastic fluctuations in both structures are
expected to be identical to order $(1/1000)^2$ so that the transition field
should not be affected by temperature, so long as the input parameters 
(e.g., $\xi_{{\bf k}}, \Delta_{{\bf k}}, \lambda$ ) do not change with $T$.  The
same cannot be said of the high field $(\sim 28T)$ triangular to centred square
lattice transition, because one has a tight packed structure and the other
not.  The expectation is that the former has lesser elastic fluctuations than
the latter, so that the transition field boundary should shift to lower values
with increasing temperature.  However, this conclusion is tempered by the fact
that the London approximation is unreliable at these high fields when vortex
cores get close to each other so that our basic result may not be that
reliable.

One interesting consequence of the orientational transition at 5T, which might
be measurable, is the change in the very low energy density of quasiparticle
states.  At least for a square lattice, Vishwanath \cite{vishwanath}
has shown that there are quasiparticle states with linear dispersion
and that there is a very small gap arising from higher
order terms in the quasiparticle velocity.  If this kind of result carries
through for a  triangular lattice, then it might be an experimental way of
observing the transition.

We have calculated here, for the first time, the actual energy of the
structure sensitive term (Section III) and have found it to be small, of
order a few degrees per vortex.  Because of this reason, the transition might
be difficult to observe, since pinning energies of larger size are
generally present \cite{attanasio},
unless the system is extremely perfect.

\begin{figure}[tb]
\psfig{file=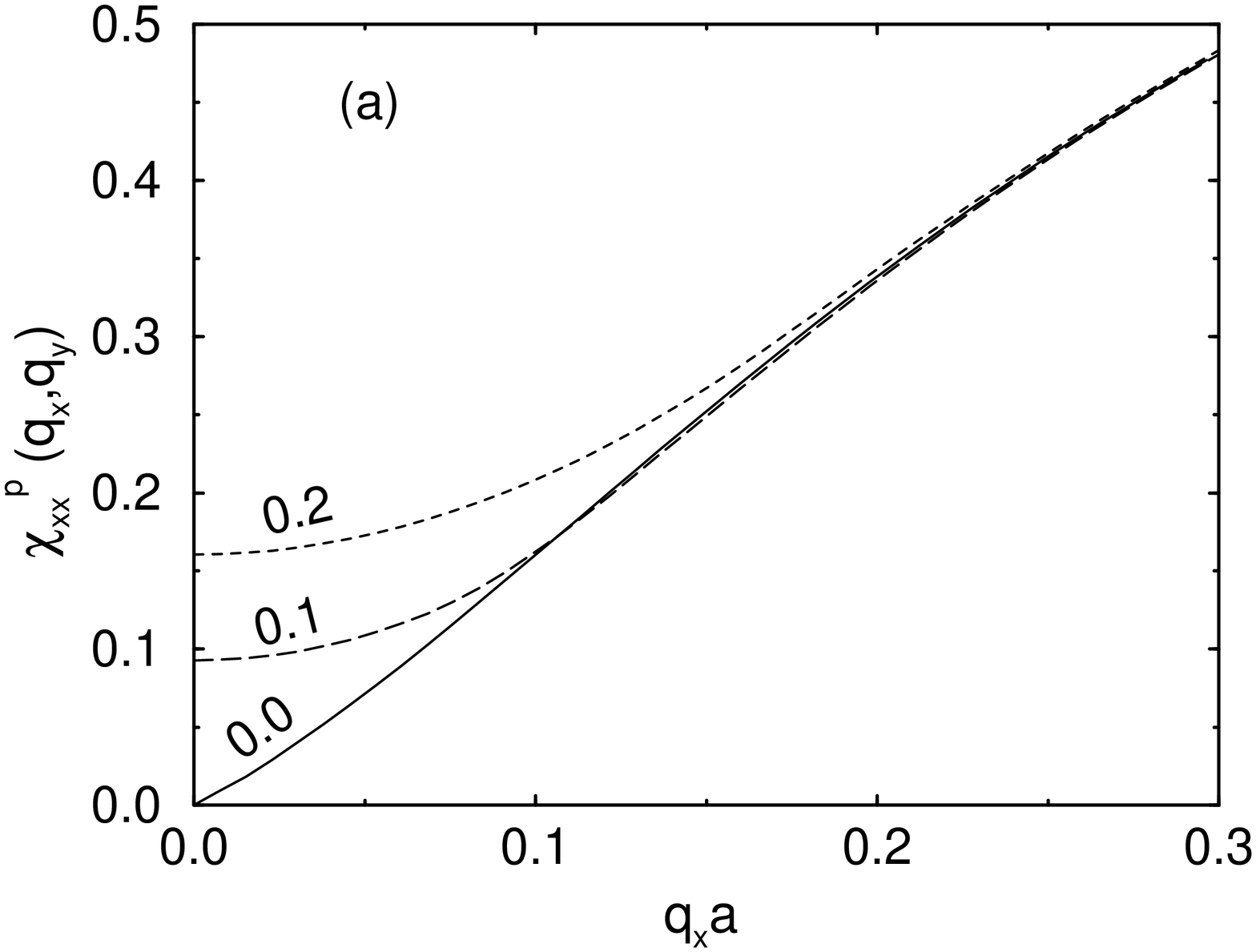,width=12cm,angle=-90}
\psfig{figure=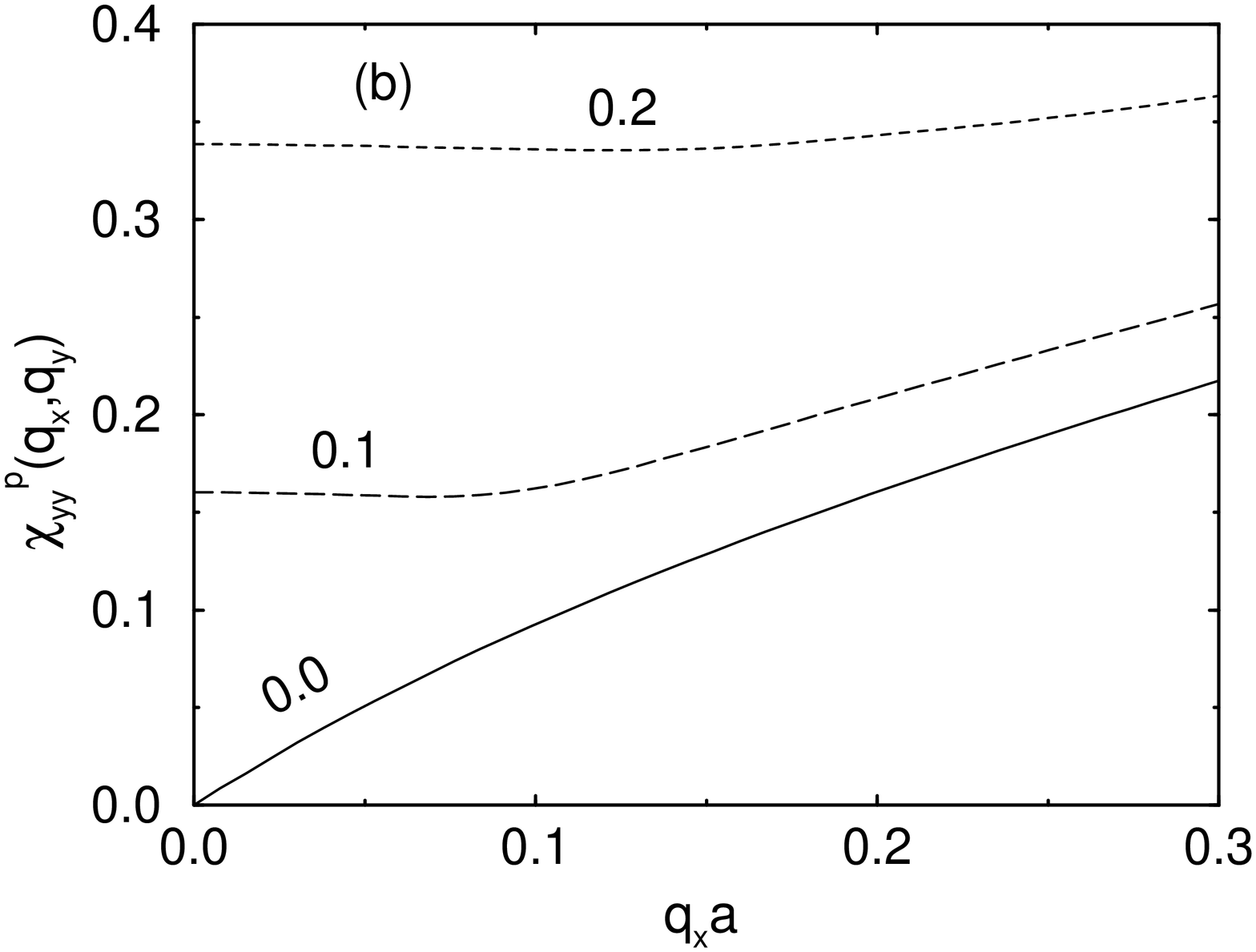,width=12cm,angle=-90}
\psfig{figure=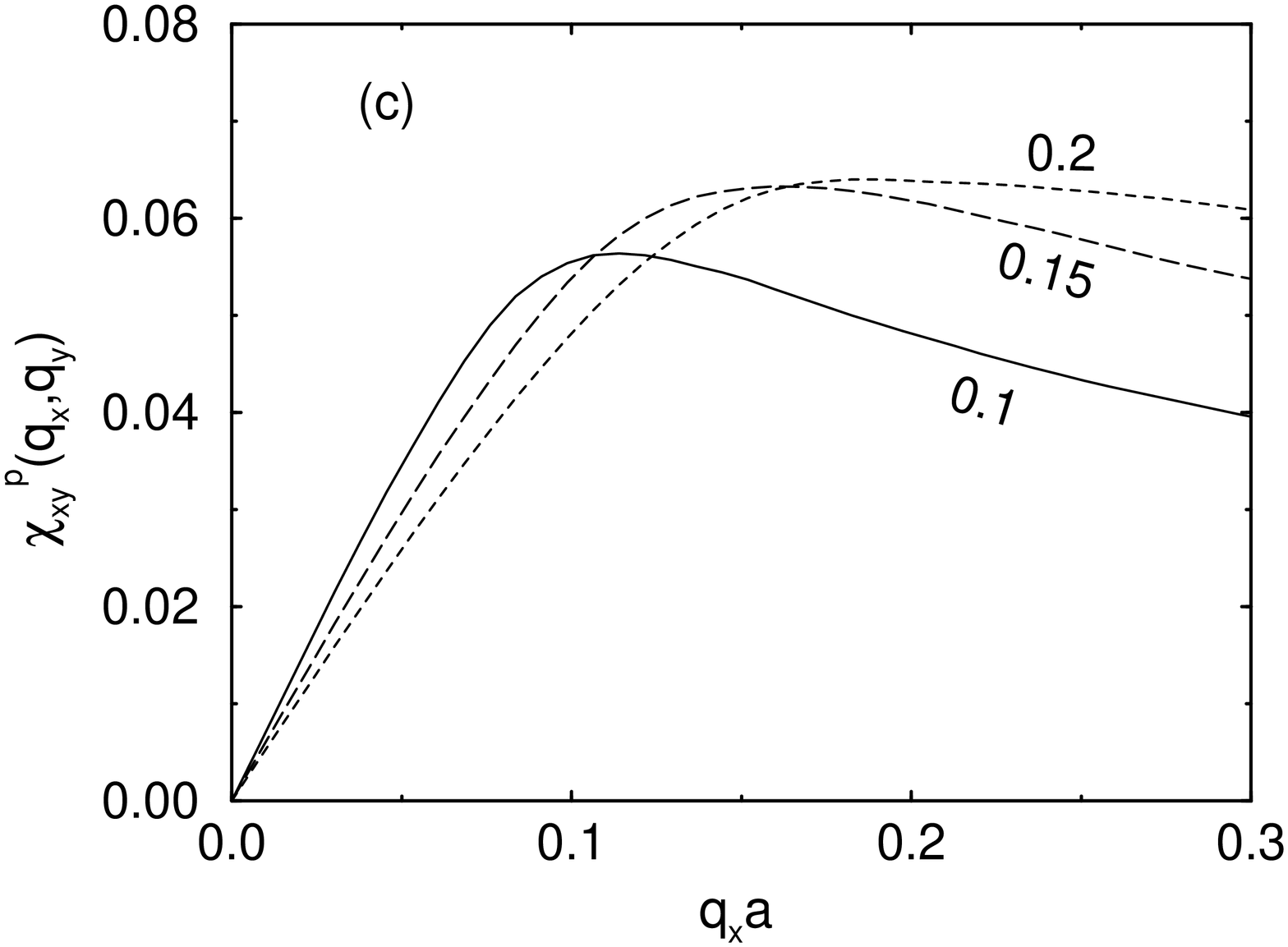,width=12cm,angle=-90}
 \caption{Dimensionless susceptibilities(a) $\chi^{{\rm
p}}_{xx}$, 
(b) $\chi^{{\rm p}}_{yy}$ and (c) $\chi^{{\rm p}}_{xy}$
(in units of $\chi_d$)
plotted against small positive $q_xa$. The
 numbers associated with each curve are the corresponding values
of $q_ya$. The
 susceptibilities for negative values of $q_x$ and $q_y$ are
related to the same
 for positive values of $q_x$ and $q_y$ by the symmetries
discussed in the text.}
\label{chiab}
\end{figure}

We have discussed in detail (Section II and III) the reason why our results
differ from those obtained earlier. Basically, it has to do with the
anisotropy of the nonlocal current current susceptibility, i.e. the fact that
$\chi^{{\rm p}}_{xx}({\bf G}) \neq \chi^{{\rm p}}_{yy}({\bf G})$.  The reason for this essentially
is that we have an anisotropic superconductor.  The $\chi^{{\rm p}}_{xx}$ and
$\chi^{{\rm p}}_{yy}$ functions are plotted in Fig.\ref{chiab}.

The question of a non triangular structure of the vortex lattice in cuprates
has attracted considerable experimental attention \cite{maprile,keimer,johnson}, 
especially since
it has become established that they are $d_{x^{2}-y^{2}}$ superconductors. 
Earlier small angle neutron scattering measurements \cite{keimer} were on highly
twinned 123 crystals, so that the observation of four fold diffraction
symmetry does not imply rectanglar /square lattice.  Moreover, the positional
order is very poor.  A more recent experiment \cite{johnson} on untwinned 123 single
crystals shows much better translational order (higher $G$ peaks are resolved
for the first time) and a triangular lattice with axes oriented along a,
distorted because of $a$-$b$  asymmetry.  The authors find no structural
transitions upto 4T with field along $c$-axis.  They however find a transition
from a triangular lattice oriented along $a$ to one oriented
along the $b$ axis at a field of about 3.8T, oriented at 33$^o$ to the
$c$ axis.  This is certainly quite different from the transition to a
triangular lattice at 45$^o$ to $a$ axis at 5T predicted by us.  As Johnson et
al point out \cite{johnson}  the observed transition could be due to the
presence of chains in 123, and the novel $ab$-$c$ anisotropy caused by it
(which may have a strong effect on many physical properties).  In order to
seriously explore our conclusions, one needs to do experiments on cuprates
without chains and ideally with tetragonal symmetry, as again mentioned by
Johnson et al \cite{johnson}.

\acknowledgements{We acknowledge support from the JNC India and the US-India
  ONR funded project No. N00014-97-0988.}

\newpage

\begin{references}
\bibitem[*]{add1} Also at Condensed Matter Theory Unit, JawaharlaNehru Centre for Advanced Scientific Research, Jakkur, Bangalore
560 064,India.
\bibitem[\dagger]{add2} Present address: Department of Physics, 104 Davey Laboratory,
  The Pennsylvania State University, University Park, PA 16802.

\bibitem{abrikosov} A. A. Abrikosov, Sov. Phys. JETP {\bf 5}, 1174 (1957).

\bibitem{essmann} U. Essmann and H. Trauble, Phys. Lett. {\bf 24A}, 526 (1967).

\bibitem{hess} H. F. Hess, R. B.  Robinson, R. C. Dynes,
      J. M. Valles, Jr., and J. V. Waszczak,
    Phys. Rev. Lett. {\bf 62}, 214 (1989).

\bibitem{h_fermion} R. N. Kleiman, C. Broholm, G. Aeppli,
         E. Bucher, N. Stucheli, D. J. Bishop, K. N. Clausen,
         K. Mortensen, J. S. Pedersen, and B. Howard,
          Phys. Rev. Lett. {\bf 69}, 3120 (1992).

\bibitem{riseman} T. M. Riseman, P. G. Kealey, E. M. Forgan,
              A. P. Mackenzie, L. M. Galvin, A. W. Tyler,
              S. L. Lee, C. Ager, D. McK. Paul, C. M. Aegerter,
              R. Cubitt, Z. Q. Mao, T. Akima, and Y. Maeno,
              Nature (London) {\bf 396}, 242 (1998); 
         P. G. Kealey, T. M. Riseman, E. M. Forgan, L. M. Galvin,
          A. P. Mackenzie, S. L. Lee, D. McK. Paul, R. Cubitt,
          D. F. Agterberg, R. Heeb, Z. Q. Mao, and Y. Maeno,
       Phys. Rev. Lett. {\bf 84}, 6094 (2000).

\bibitem{maprile} I. Maggio-Aprile, Ch. Renner, A. Erb, E. Walker,
        and $\phi$. Fischer, Phys. Rev. Lett. {\bf 75}, 2754 (1995).            

\bibitem{keimer} B. Keimer, W. Y. Shih, R. W. Erwin, J. W. Lynn,
     F. Dogan, and I. A. Aksay, Phys. Rev. Lett. {\bf 73}, 3459 (1994).

\bibitem{johnson} S. T. Johnson, E. M. Forgan, S. H. Lloyd,
     C. M. Aegerter, S. L. Lee, R. Cubitt, P. G. Kealey, C. Ager,
     S. Tajima, A. Rykov, and D. McK. Paul,
    Phys. Rev. Lett. {\bf 82}, 2792 (1999).

\bibitem{ichioka} M. Ichioka, N. Hayashi, N. Enomoto, and K.
    Machida, Phys. Rev. B {\bf 53}, 15 316 (1996).

\bibitem{maki} H. Won and K. Maki, Phys. Rev. B {\bf 53}, 5927 (1996).

\bibitem{affleck} I. Affleck, M. Franz, and M. H. S. Amin,
    Phys. Rev. B {\bf 55}, R704 (1997).

\bibitem{franz} M. Franz, I. Affleck, and M. H. S. Amin,
    Phys. Rev. Lett. {\bf 79}, 1555 (1997).

\bibitem{amin} M. H. S. Amin, I. Affleck, and M. Franz,
    Phys. Rev. B {\bf 58}, 5848 (1998).

\bibitem{blatter} G. Blatter, M. V. Feigel'man, V. B. Geshkenbein,
      A. I. Larkin, and V. M. Vinokur, Rev. Mod. Phys., {\bf 66}, 1125 (1994); 
      A. Schilling, R. A> Fisher, N. E. Phillips, U. Welp,
   D. Dasgupta, W. K. Kwok, and G. W. Crabtree,
    Nature (London) {\bf 382}, 791 (1996); 
      R. J. Drost, C. J. van der Beek, J. A. Heijn,
      M. Konczykowski, and P. H. Kes, Phys. Rev. B {\bf 58}, R615 (1998); 
      S. L. Lee, C. M. Aegerter, S. H. Lloyd, E. M. Forgan,
      C. Ager, M. B. Hunt, H. Keller, I. M. Savic, R. Cubitt,
      G. Wirth, K. Kadowaki, and N. Koshizuka, 
      Phys. Rev. Lett. {\bf 81}, 5209 (1998); 
      M. J. W. Dodgson, V. B. Geshkenbein, and G. Blatter, 
      Phys. Rev. Lett. {\bf 83}, 5358 (1999); 
      C. J. van der Beek, M. Konczykowski, R. J. Drost, P. H. Kes,
      N. Chikumoto, S. Bouffard,
      Phys. Rev. B {\bf 61}, 4259 (2000); 
   M. Muller, D. A. Gorokhov, and
      G. Blatter, Phys. Rev. B {\bf 64}, 134523 (2001).

\bibitem{expts} W. N. Hardy, D. A. Bonn, D. C. Morgan, R. Liang,
     and K. Zhang, Phys. Rev. Lett. {\bf 70}, 3999 (1993);
    D. A. Wollman, D. J. van Harlingen, W. C. Lee, D. M. Ginsberg,
     and A. J. Leggett, Phys. Rev. Lett. {\bf 71}, 2134 (1993);
    C. C. Tsuei, J. R. Kirtley, C. C. Chi, L. S. Yu-Jajnes,
     A. Gupta, T. Shaw, J. Z. Sun, and M. B. Ketchen,
    Phys. Rev. Lett. {\bf 73}, 593 (1994).

\bibitem{norman2} H. Ding, M. R. Norman, J. C. Campuzano, M. Randeria,
     A. F. Bellman, T. Yokoya, T. Takahasi, T. Mochiku, and
     K. Kadowaki, Phys. Rev. B {\bf 54}, R9678 (1996).

\bibitem{bonn} D. A. Bonn, P. Dosanjh, R. Liang, and W. N. Hardy,
    Phys. Rev. Lett. {\bf 68}, 2390 (1992); 
    K. Krishana, J. M. Harris, and N. P. Ong, 
     Phys. Rev. Lett. {\bf 75}, 3529 (1995);
  K. Krishana, N. P. Ong, Y. Zhang, Z. A. Xu, R. Gagnon, and
    L. Taillefer, Phys. Rev. lett. {\bf 82}, 5108 (1999);
   A. Hosseini, R. Harris, S. Kamal, P. Dosanih, J. Preston,
  R. Liang, W. N. Hardy, and D. A. Bonn, Phys. Rev. B {\bf 60}, 1349 (1999).

\bibitem{norman1} M. R. Norman, M. Randeria, H. Ding, and J. C.
Campuzano, Phys. Rev. B {\bf 52}, 615 (1995).


\bibitem{moler} K. A. Moler, D. J. Baar, J. S. Urbach, R. Liang,
     W. N. Hardy, and A. Kapitulnik, Phys. Rev. Lett. {\bf 73}, 2744 (1994);
   B. Revaz, J.-Y. Genoud, A. Junod, K. Neumaier, A. Erb,
     and E. Walker, Phys. Rev. Lett. {\bf 80}, 3364 (1998);
   D. A. Wright, J. P. Emerson, B. F. Woodfield, J. E. Gordon,
  R. A. Fisher, and N. E. Phillips, Phys. Rev. Lett. {\bf 82}, 1550 (1999).

\bibitem{krishana} K. Krishana, N. P. Ong, Q. Li, G. D. Gu,
  and N. Koshizuka, Science {\bf 277}, 83 (1997); 
H. Aubin, K. Behnia, S. Ooi, and T. Tamegai,
  Phys. Rev. Lett. {\bf 82}, 624 (1999);
B. Zeini, A. Freimuth, B. Buchner, R. Gross, A. P. Kampf, 
  M. Klaser, and H. M.-Vogt, Phys. Rev. Lett. {\bf 82}, 2175 (1999);
M. Chiao, R. W. Hill, C. Lupien, B. PopiC, R. Gagnon,
 and L. Taillefer, Phys. Rev. Lett. {\bf 82}, 2943 (1999);
N.P. Ong, K. Krishana, Y. Zhang, and Z.A. Xu,
     in {\it Physics and Chemistry of Transition Metal Oxides},
  edited by H. Fukuyama and N. Nagaosa (Springer-Verlag, Berlin, 1999), p.202; 
Y. Ando, J. Takeya, Y. Abe, K. Nakamura, and A. Kapitulnik,
   Phys. Rev. B {\bf 62}, 626 (2000).


\bibitem{sonier} J.E. Sonier, J.H. Brewer, R.E. Kiefl, G.D. Morris,
     D.A. Bonn, J. Chakhalian, R.H. Heffner, W.N. Hardy, and R. Liang,
    Phys. Rev. Lett. {\bf 83}, 4156 (1999).

\bibitem{leggett} I. Kosztin and A. J. Leggett, 
    Phys. Rev. Lett. {\bf 79}, 135 (1997).

\bibitem{prrm} A. Paramekanti, M. Randeria, T. V. Ramakrishnan,
    and S. S. Mandal, Phys. Rev. B {\bf 62}, 6786 (2000).

\bibitem{bonn2} S. Kamal, R. Liang, A. Hosseini, D. A. Bonn,
     and W. N. Hardy, Phys. Rev. B {\bf 58}, R8933 (1998).

\bibitem{dagotto} E. Dagotto, Rev. Mod. Phys., {\bf 66}, 763 (1994).

\bibitem{volovik} G. E. Volovik, JETP Lett. {\bf 58}, 469 (1993).

\bibitem{tesanovic} M. Franz, Z. Tesanovoc, Phys. Rev. Lett. {\bf 84},
            554 (2000); O. Vafek, A. Melikyan, M. Franz, and
      Z. Tesanovic, Phys. Rev. B {\bf 63}, 134509 (2001).

\bibitem{halperin} L. Marinelli, B. I. Halperin, and S. H. Simon,
        Phys. Rev. B {\bf 62}, 3488 (2000).

\bibitem{vishwanath} A. Vishwanth, arXiv:con-mat/0104213.

\bibitem{attanasio} See for example: C. Attanasio, L. Maritato, 
     C. Coccorese, S. L. Prischepa, A. N. Lykov, and M. Salvato, 
   Trans. Appl. Supercond. {\bf 5}, 1359 (1995).
 
\end{references}
\end{document}